\documentclass[12pt]{article}

\pdfoutput=1

\usepackage{graphicx}
\usepackage{youngtab}
\usepackage[centertags]{amsmath}
\usepackage{amssymb}
\usepackage{amsthm}
\usepackage{amsfonts}
\usepackage{ccaption}
\usepackage[usenames]{color}
\usepackage{mathrsfs}

\usepackage[
      colorlinks=true,
      linkcolor=blue,
      urlcolor=blue,
      filecolor=black,
      citecolor=red,
      pdfstartview=FitV,
      pdftitle={},
        pdfauthor={Mukund Rangamani},
        pdfsubject={},
        pdfkeywords={},
        pdfpagemode=None,
        bookmarksopen=true
      ]{hyperref}

\usepackage{rotating}

\makeatletter
\@addtoreset{equation}{section}

\makeatletter
\renewcommand\section{\@startsection {section}{1}{\z@}%
                                   {-3.5ex \@plus -1ex \@minus -.2ex}
                                   {2.3ex \@plus.2ex}%
                                   {\normalfont\large\bfseries}}
\renewcommand\subsection{\@startsection{subsection}{2}{\z@}%
                                     {-3.25ex\@plus -1ex \@minus -.2ex}%
                                     {1.5ex \@plus .2ex}%
                                     {\normalfont\bfseries}}

  \captionnamefont{\bfseries}
  \captiontitlefont{\small\sffamily}
  \captiondelim{: }
  \hangcaption

\def\sec#1{\S\ref{#1}}
\def\fig#1{Figure\,\ref{#1}}
\def\req#1{(\ref{#1})}
\def\App#1{Appendix \ref{#1}}

\def\thus{\Longrightarrow}

\def\CB{{\cal B}}
\def\CC{{ \cal C }}
\def\CD{{ \cal D }}

\def\CL{{\cal L}}

\def\CN{{\cal N}}
\def\CO{{\cal O}}

\def\CS{{\cal S}}

\def\CV{{\cal V}}

\def\ZZ{\mathbb{Z}}

\def\R{{\bf R}}
\def\Sp{{\bf S}}

\def\vev#1{\langle\, #1 \, \rangle}

\def\ord#1{\CO\left( #1 \right)}

\def\Tr#1{{\rm Tr}\left(#1\right)}

\definecolor{purple}{rgb}{0.8,0.3,0.5}
\definecolor{green}{rgb}{0.1,0.8,0.2}

\def\AdS#1{AdS$_{#1}$}
\def\SAdS#1{Schwarzschild-AdS$_{#1}$}
\def\ESU#1{ESU$_{#1}$}
\def\scri{\mathscr I}

\title{{\bf \Large Conformal field theories in anti-de Sitter space}}

\author{\normalsize
Ofer Aharony$^a$\footnote{Ofer.Aharony@weizmann.ac.il}\ ,\  Donald Marolf$^{\,b}$\footnote{marolf@physics.ucsb.edu}\ ,  Mukund Rangamani$^{c,d}$\footnote{mukund.rangamani@durham.ac.uk} \\
\small \sl $^a$  Department of Particle Physics and Astrophysics,
\\[-1.5mm] \small \sl Weizmann Institute of Science, Rehovot 76100, Israel. \\
\small \sl $^b$  Physics Department, UCSB, Santa Barbara, CA 93106, USA. \\
\small \sl $^c$  Centre for Particle Theory \& Department of
Mathematical Sciences,
\\[-1.5mm]
\small \sl Science Laboratories, South Road, Durham DH1 3LE, UK. \\
\small \sl $^d$ Albert Einstein Minerva Center, Weizmann Institute of
Science, Rehovot 76100, Israel. \\
}

\begin{document}

\setlength{\baselineskip}{16pt}
\begin{titlepage}
\maketitle
\begin{picture}(0,0)(0,0)
\put(340,320){DCPT-10/37}
\put(340,303){WIS/16/10-NOV-DPPA}
\end{picture}
\vspace{-36pt}

\begin{abstract}
In this paper we discuss the dynamics of conformal field theories on anti-de Sitter
space, focussing on the special case of the ${\cal N}=4$ supersymmetric Yang-Mills
theory on \AdS{4}. We argue that the choice of boundary conditions, in particular for
the gauge field, has a large effect on the dynamics. For example, for weak coupling,
one of two natural choices of boundary conditions for the gauge field leads to a large $N$ deconfinement phase transition
as a function of the temperature, while the other does not. For boundary conditions that
preserve supersymmetry, the strong coupling dynamics can be analyzed using S-duality
(relevant for $g_{YM} \gg 1$), utilizing results of Gaiotto and Witten,
as well as by using the AdS/CFT correspondence (relevant for
large $N$ and large 't Hooft coupling). We argue that some very specific choices of
boundary conditions lead to a simple dual gravitational description for this theory,
while for most choices the gravitational dual is not known. In the cases where the
gravitational dual is known, we discuss the phase structure at large 't Hooft coupling.
 \end{abstract}
\thispagestyle{empty}
\setcounter{page}{0}
\end{titlepage}

\renewcommand{\thefootnote}{\arabic{footnote}}


\tableofcontents

\section{Introduction and summary of results}
\label{s:intro}

Quantum fields in curved spacetime have been the source of many interesting surprises and useful theoretical insight.  Many investigations of quantum dynamics in curved backgrounds have focussed on  fields in spacetimes with non-trivial causal structure, such as black holes  or cosmological backgrounds. Nevertheless, it is also interesting to consider fields in causally trivial backgrounds, with the hope of gaining not only some insights  into the response of the field theory to background curvature, but also to ascertain the non-trivial constraints imposed by the background on the quantum dynamics.

In this paper, we focus on an important example of this type, the dynamics of conformally invariant quantum field theories  (CFTs) on global Anti de Sitter (AdS) spacetime. The interest in this background is manifold. For one, AdS is a conformally flat geometry and is a  maximally symmetric spacetime. Recall that in $d$ dimensions, \AdS{d} has as its isometry group $SO(d-1,2)$. This isometry algebra is as large as the Poincar\'e algebra, and this makes \AdS{} a natural arena for investigating the behaviour of $d$-dimensional CFTs; the latter have a global $SO(d,2)$ symmetry which includes the AdS isometries. The AdS geometry is not globally hyperbolic, and in fact has a time-like future null infinity $\scri^+$. This implies that one has to prescribe boundary conditions in order to completely specify any dynamical system on this background, be it classical or quantum. In particular, this fact allows one to investigate in a covariant manner the effect of boundary conditions on the dynamics of CFTs.

In fact, in the early days of supergravity it was realized that there are interesting boundary conditions that can play a role in the dynamics of fields in AdS spacetime \cite{Breitenlohner:1982jf}. The interest in those days was in understanding gauged supergravity theories; our main consideration is for quantum dynamics of CFTs on a fixed (non-dynamical) AdS background. It was also realized in \cite{Callan:1989em} that the AdS spacetime provides a geometric and uniquely symmetric (by virtue of its maximal symmetry) infra-red cut-off which makes it useful to study dynamics of field theories which require an infra-red regulator. The discussion of \cite{Callan:1989em} was incomplete in that it did not account for the most general set of boundary conditions available, which as we shall see play a rather crucial role in the dynamics. Of particular interest to us will be the dynamics of strongly coupled conformal gauge theories in AdS backgrounds. This has two important physical applications, which we now describe.

Let us first consider the case where we have a strongly coupled CFT on \AdS{d} whose dynamics we understand, and then turn on the gravitational interaction weakly. It is a well known fact that any consistent gravitational theory on AdS spacetime has a dual holographic avatar from the AdS/CFT correspondence. Specifically, if it can be embedded into a consistent background, the dynamics of weak gravity together with the strongly coupled CFT on the \AdS{d} background, can be described by a dual CFT$_{d-1}$ which lives on a space isomorphic to the boundary of \AdS{d}. Usually in the context of the AdS/CFT correspondence, the couplings of the matter (CFT$_d$) degrees of freedom are commensurate with the gravitational coupling. There is however no reason not to have a hierarchical separation of the couplings, and the models we focus on correspond to the extreme limit where gravitational interactions have been switched off (which is always a consistent setup in itself). Understanding the dual field theory as we slowly turn on the gravitational interaction can provide interesting insights into the AdS/CFT correspondence and into the dynamics of field theories in AdS.

On the other hand, one can start with a $d$-dimensional conformal field theory which is known to have a holographic dual in terms of gravity in a higher dimensional \AdS{d+1} spacetime. Consider for instance the well known duality between $ {\cal N} =4$ Supersymmetric Yang-Mills (SYM) in four dimensions and  type IIB string theory on \AdS{5} $\times \ {\bf S}^5$. One can consider the dynamics of $ {\cal N}=4$ SYM on any background geometry; in particular \AdS{4} is an appropriate (and for reasons mentioned earlier a particularly interesting) candidate.
To understand the holographic dual of ${\cal N} =4$ SYM on \AdS{4} one would need to construct asymptotically locally \AdS{5}$\times \ {\bf S}^5$ geometries whose boundary is \AdS{4}. While a class of such geometries is easy to construct, using the fact that \AdS{5} can be written in an \AdS{4} foliation, it turns out that the precise answer is much more subtle than naively anticipated.

To appreciate the subtlety let us return to the issue of boundary conditions to which we have previously alluded. If there was a unique choice of boundary conditions for the ${\cal N}=4$ SYM fields on \AdS{4}, then one would indeed be tempted to look for ``the holographic dual'' of this system in terms of the aforementioned \AdS{4} foliations of \AdS{5}. However, the choice of available boundary conditions is quite rich and large, and different boundary conditions lead to different theories and different gravitational duals, lending this system much of its intricacy. It therefore behooves us to understand both the choices of available boundary conditions and their implications for the dynamics of the system.

The richness accorded by the boundary conditions can be understood by recalling that for scalar fields in \AdS{d} in a certain mass range, close to the Breitenlohner-Freedman bound,  one is allowed two distinct choices of local boundary conditions that are linear in the field and consistent with the AdS symmetries. In the AdS/CFT context, these two choices lead to different dual conformal field theories. This fact was already appreciated in the seminal work of Breitenlohner and Freedman \cite{Breitenlohner:1982jf}, and was clarified in the AdS/CFT context in \cite{Klebanov:1999tb}. In fact, it turns out that one can also pick from a choice of boundary conditions for massless vector fields in \AdS{4} \cite{Breitenlohner:1982jf}; this was used to understand the S-duality transformation properties of Abelian gauge fields in \AdS{4} in \cite{Witten:2003ya}. A more comprehensive discussion was subsequently provided in \cite{Marolf:2006nd}, based on the linearized analysis of \cite{Ishibashi:2004wx}. Since almost all known conformal theories in four spacetime dimensions are gauge theories, these general boundary conditions are of relevance in the study of interacting CFTs on \AdS{4} backgrounds.

Focussing on gauge theories in four dimensions, it transpires that one can impose at least two distinct boundary conditions which we call Dirichlet/standard/electric or
Neumann/modified/magnetic, respectively \cite{Breitenlohner:1982jf}. For the Dirichlet choice, one fixes the boundary value of the gauge field, while with the Neumann boundary condition, it is the normal component of the field strength to the boundary that gets fixed.  For an Abelian gauge theory in AdS these two boundary conditions are S-dual \cite{Witten:2003ya}. However, for non-Abelian theories, despite this innocuous sounding distinction, these two boundary conditions operate quite differently. A crucial difference between the two choices is that with the Dirichlet boundary conditions, one is allowed to consider electrically charged states in the AdS geometry; the flux lines from the charge simply go off to the boundary. However, with the Neumann boundary conditions, one has a Gauss' law constraint operating on the charged states; electrically charged states are forbidden, and instead one is allowed magnetically charged states.

A key point is that in perturbation theory, one usually has light electric states, but heavy magnetic states. This implies that for weakly coupled field theories, one should expect to see only a few light degrees of freedom for Neumann boundary conditions.
For weakly coupled field theories with large gauge groups, the distinction between the two boundary conditions is thus quite striking: for the Dirichlet boundary conditions one has ${\cal O}(\text{dim}({\mathfrak g}))$ light states for a gauge theory with gauge group $G$,\footnote{Lie groups will be denoted with upper-case Latin alphabet, $G$, $H$ etc., and the corresponding Lie algebra in lower-case gothic font, ${\mathfrak g}, {\mathfrak h}$ etc..}  while the Neumann boundary conditions only allow $ {\cal O}(1)$ light states.

If the Dirichlet and Neumann boundary conditions were to be exchanged under S-duality, as in the Abelian theory, then one would have an interesting scenario. For, starting with the Dirichlet boundary conditions for the gauge theory, with $\text{dim}({\mathfrak g}) \gg 1$, one would have a large number of electric states, which under the S-duality map would morph into a few magnetic states due to the Neumann boundary condition. Of course, this statement involves
an unjustified extrapolation from weak coupling to strong coupling, and it is also suspicious since a pure gauge theory does not have S-duality; one would need to consider a gauge theory with additional matter degrees of freedom, such as for instance the $ {\cal N} =4$ SYM, which does exhibit S-duality. In this case one has to worry about the boundary conditions on the matter fields as well, in order to make any statements about S-duality and exchange of boundary conditions. In any case, as we shall review below, the Dirichlet and Neumann boundary condition are not S-dual in non-Abelian gauge theories, thereby providing a simple resolution to the disparity in the low energy spectrum.

Before getting to the issue of the S-duals of various boundary conditions, it is also worthwhile to pause to ask whether one can use the holographic gauge/gravity correspondence to study field theories on \AdS{d} backgrounds. For superconformal field theories such as $ {\cal N} = 4$ SYM, the answer must clearly be yes, since we know that the theory on $ {\bf R}^{3,1}$ or on $ {\bf R} \times {\bf S}^3$ does indeed have a holographic dual in terms of type IIB string theory on \AdS{5} $\times\  {\bf S}^5$. In general one can consider the CFT on any curved manifold $ {\cal B}_4$, which then requires that the bulk spacetime includes an asymptotically locally AdS geometry $ {\cal M}_5$ whose boundary is in the same conformal class as $ {\cal B}_4$. The simplest such geometries which one can write down are \AdS{4} foliations of \AdS{5} (used for instance in the brane-world context in \cite{Karch:2000ct}, see \fig{f:adsinads}).\footnote{To be specific, as we discuss below, one would need to take an appropriate quotient of these geometries to obtain a single copy of the \AdS{4} spacetime on the boundary. Without such a quotient, these spacetimes are more relevant for the transparent boundary conditions on the double-\AdS{4} boundary, as discussed in \cite{Hubeny:2009rc}.}

There is however an interesting puzzle associated with the Dirichlet boundary conditions for large $N$ gauge theories from the gravity viewpoint. We know that the free theory with Dirichlet boundary conditions has a large number of light excitations, since the electrically charged adjoint fields of $G$ are admissible operators for these boundary conditions. If these states remained light as we increased the 't Hooft coupling constant of the theory, then   one expects that the holographic dual of $ {\cal N} =4$ SYM at strong coupling will correspond to a geometry with $ {\cal O} (\text{dim}({\mathfrak g}))$ states. Note that these states should live in all of AdS and not just in the interior as in black hole backgrounds; for instance, with these boundary conditions the group $G$ is a global symmetry, and there should be $\text{dim}({\mathfrak g})$ gauge fields in the bulk whose boundary conditions determine the sources for the global symmetry currents. In the large $N$ limit this bulk gauge group is strongly coupled (at least in the limit of large $g_s N$ where one expects the holographic dual to be under control), so it seems unlikely that any weakly coupled description of the bulk physics exists for these boundary conditions (unless there is some phase transition at a finite value of the 't Hooft coupling in which almost all the light states disappear); the naive holographic dual mentioned above cannot be valid.

It turns out that in order to understand the strongly coupled theory in light of both the issues with S-duality and the puzzles associated with the holographic description, one has to go back to the drawing board and carefully disentangle the way the boundary conditions are imposed. For a theory like $ {\cal N} =4$ SYM one has a large number of boundary conditions that can be imposed on the boundary of \AdS{4}, and not all of them are expected to be tractable. We need to identify those that can be followed from weak to strong coupling, using either the electromagnetic S-duality of the theory or holographic methods. The class of boundary conditions which best lend themselves to such an analysis are the supersymmetric boundary conditions preserving sixteen supercharges. It turns out that there is a vast number of such boundary conditions. Fortunately, these have been discussed in detail for a closely related problem by Gaiotto and Witten \cite{Gaiotto:2008sa}, who have also described the role of S-duality for these same boundary conditions in \cite{Gaiotto:2008ak}. The analysis in \cite{Gaiotto:2008sa,Gaiotto:2008ak} was for the field theory on the half-space $ {\bf R}^{2,1} \times {\bf R}_+$, which can be realized by D3-branes ending on other branes, but this geometry is conformal to \AdS{4}; we will exploit this feature to port the results of \cite{Gaiotto:2008sa, Gaiotto:2008ak} to the situation at hand. As we will discuss, the choice of a boundary condition for
the theory on the half-space maps to a choice of supersymmetric vacuum for the theory on \AdS{4}.

 One main feature of the supersymmetric boundary conditions for $ {\cal N} = 4$ SYM on \AdS{4} with $G=SU(N)$ is that the simplest Dirichlet and Neumann boundary conditions are no longer S-dual to each other \cite{Gaiotto:2008sa,Gaiotto:2008ak}. In fact, the S-dual of the simplest Neumann boundary condition does involve a Dirichlet-type boundary condition, but it is one in which the gauge group $G$ turns out to be broken completely. This is easy to understand in the D-brane language: the Neumann boundary conditions may be realized by requiring $N$ D3-branes to end on  a single NS5-brane (which cannot accept fundamental string charge, and so forces the total electric charge on the D3-branes to vanish).  Thus, the S-dual involves D3-branes ending on a single D5-brane.  But since the world-volume flux on a single D5-brane is Abelian, it is not possible for a non-Abelian configuration of D3-branes to end on a single D5-brane.  Rather, one must `Higgs' the D3-brane world-volume theory appropriately so that its gauge group is completely broken.
 These boundary conditions are obtained by prescribing a class of allowed singularities for the scalars of ${\cal N} = 4$ SYM on the boundary of the half-space, which break the gauge symmetry, and via a conformal transformation lead to vacuum expectation values (VEVs) for the scalars on \AdS{4}, and thus to masses (and modified conformal weights) for the corresponding gauge fields in \AdS{4}.

The simplest Dirichlet boundary condition for $ {\cal N} =4$ SYM with $G=SU(N)$ (that does not break the gauge symmetry) actually arises from $N$ D3-branes ending on a commensurate number of D5-branes. Its S-dual, corresponding to $N$ D3-branes ending on $N$ NS5-branes, has a non-trivial $2+1$ dimensional CFT living on the boundary, coupled to the dynamics of the ${\cal N} =4$ SYM degrees of freedom via the boundary conditions. In this case the theory on \AdS{4} has an $SU(N)$ global symmetry, implying that a
potential gravitational dual must include $SU(N)$ gauge fields, so as mentioned above it cannot be described by
a background with a fixed spectrum in the large $N$ limit, and it seems that due to the large number of fields the dual is not weakly coupled (at least near the boundary). Note that even the Neumann boundary condition described above presents a challenge for finding a gravitational dual, since it requires understanding the near-horizon geometry of $N$ D3-branes ending on a single NS5-brane, which is a challenging proposition; however, in this case it is possible that a smooth
gravitational dual could exist (though it is not yet known).

This raises the question of whether we can construct any examples of boundary conditions for ${\cal N} =4$ SYM on \AdS{4} which can be usefully analyzed at large 't Hooft coupling using holographic techniques. To this end we investigate a more general class of boundary conditions described in \cite{Gaiotto:2008sa, Gaiotto:2008ak},  involving orientifold/orbifold fixed planes. We argue that there are specific classes of such examples, whose string theory dual involves $O5$-planes or orbifold loci whose world-volumes are \AdS{4} $\times \;{\bf S}^2$ $\subset$ \AdS{5} $\times \;{\bf S}^5$ in type IIB string theory. With these more general boundary conditions one breaks the gauge group $G$ down to a subgroup $H$, with Neumann boundary conditions imposed on $H$-valued gauge fields (and Dirichlet on the others). For sufficiently large $H \subset G$ the theory has only a few light excitations. For such boundary conditions we find holographic duals which are weakly curved $\ZZ_2$ quotients of type IIB string theory on \AdS{5} $\times \;{\bf S}^5$  (for $G = SU(N)$), and the S-duals have similar properties.

The outline of this paper is as follows. We begin in \sec{s:qftads} with a brief summary of the boundary conditions allowed for free quantum fields in \AdS{d}.  Since our primary focus is on four dimensional gauge theories, we describe the basic issues involved in the choice of boundary conditions for massless vector fields in \sec{s:gaugeads4}, which then sets the stage for our analysis of free $ {\cal N} =4$ SYM on \AdS{4} in \sec{s:n4ads}.
In particular, we compute the spectrum  and the finite temperature partition sums of the free theory on \AdS{4} with different choices of boundary conditions. One physical result we derive is that the free theory with Neumann boundary conditions undergoes a large $N$ Hagedorn transition on \AdS{4}. The precise numerical value for the critical temperature depends on the choice of boundary conditions for the scalar fields, but it is clearly set by the \AdS{4} length scale, $\ell_4$. In \sec{s:strongN4} we take preliminary steps to understanding the strongly coupled theory; we are quickly led to a more comprehensive discussion of the boundary conditions, and this we undertake in \sec{s:bc2}. The simplest class of supersymmetric boundary conditions will turn out to be hard to understand holographically. As a result, in  \sec{s:bcquotients} we will describe the supersymmetric boundary conditions involving quotients, some of which turn out to admit holographic duals which we can use to address the strong coupling phase structure of the theory. We end in \sec{s:discuss} with a discussion of open issues. Two appendices contain some technical details.

\section{Free quantum fields on \AdS{d}}
\label{s:qftads}

Let us begin with a consideration of free quantum fields in \AdS{d}. A well known fact about \AdS{} spacetimes is that they possess a time-like future null infinity $\scri^+$, which implies that these spacetimes are not globally hyperbolic. This means that to have a well-posed Cauchy problem one needs to prescribe sensible boundary conditions on the time-like boundary, which for global \AdS{d} spacetimes is the Einstein Static Universe, \ESU{d-1} $\equiv \R \times \Sp^{d-2}$.

From early investigations on the subject \cite{Breitenlohner:1982jf} it was clear that one can have non-trivial choices of boundary conditions to impose. We will first recall some of the basic facts regarding such boundary conditions for fields up to spin $1$, since we will be primarily interested in quantum fields without gravity in asymptotically \AdS{d} backgrounds. See \cite{Ishibashi:2004wx,Compere:2008us,Amsel:2009rr} for a discussion of higher spins.  To describe the boundary conditions it is useful to fix a metric on the spacetime. We find it convenient to work with global coordinates on \AdS{d}:
\begin{equation}
ds^2 = -(1+\frac{r^2}{\ell_d^2})\, dt^2 + \frac{dr^2}{1+\frac{r^2}{\ell_d^2}} + r^2\,d\Omega_{d-2}^2\,.
\label{global2}
\end{equation}	
In these coordinates the conformal boundary \ESU{d-1} of \AdS{d} is obtained as
$r \to \infty$. We will often use intuition from the AdS/CFT correspondence and discuss the interpretation of our boundary conditions from the point of view of a hypothetical dual CFT living on \ESU{d-1}; \footnote{When gravity in \AdS{d} is non-dynamical, we do not expect to have a holographic dual in terms of a CFT$_{d-1}$. Formally one can still view the spectral data of fields in \AdS{d} in terms of representations of the isometry group $SO(d-1,2)$, which is the conformal group for a $(d-1)$-dimensional CFT. In some cases it may be possible
to really have a physical dual conformal field theory, if the quantum field theory on AdS space can be part of a consistent theory of quantum gravity, but we will not discuss this possibility here.} independently of this, in certain cases the boundary conditions will lead to having physical degrees of freedom located on the boundary \ESU{d-1}.

Another useful fact to keep in mind is that \AdS{d} itself, being conformally flat, can be mapped into a space conformal to \ESU{d} via a simple coordinate change $r = \ell_d\, \cot (\theta)$, which leads to the metric
\begin{equation}
ds^2 = \frac{1}{\sin^2(\theta)}\left[- dt^2 +\ell_d^2 \left(  d\theta^2+ \cos^2 (\theta) \; d\Omega_{d-2}^2\right) \right],
\label{globalesu}
\end{equation}	
where now the boundary is located at $\theta = 0$. It is important to note that \ESU{d} itself is a double-cover of \AdS{d}; the two copies of the \AdS{d}, $-\pi/2 < \theta < 0$ and $0 < \theta < \pi/2$, are joined across their boundaries, which sit at the equator $\theta = 0$ of the \ESU{d}.

Let us review the local linear boundary conditions that are admissible for fields of spin $s \le 1$, and which are invariant under $SO(d-1,2)$. Details of how these boundary conditions arise are compiled in \App{s:bcreview} for completeness.

\paragraph{Scalars ($s =0$):} For scalar fields of mass $m$, with a conformal coupling to the background curvature, the allowed boundary conditions depend on the effective mass $m_\text{eff}^2  = m^2 + m_c^2$, where we define the conformal mass $m_c^2 =  -\frac{d(d-2)}{4\,\ell_d^2}$. The two independent solutions to the scalar field wave equation behaves at large $r$ as
\begin{equation}
\phi(r,x) \to A(x)\,  r^{-\Delta_- } + B(x) \, r^{-\Delta_+} \ , \qquad x = \{t,\Omega_{d-2}\}
\label{}
\end{equation}	
with
\begin{equation}
\Delta_\pm = \frac{d-1}{2}  \pm \sqrt{\frac{(d-1)^2}{4} + m_\text{eff}^2 \, \ell_{d}^2} \ \ .
\label{}
\end{equation}	
 In the range $m_{BF}^2 \le m_\text{eff}^2 \le m_{BF}^2 +\ell_d^{-2}$, with $m_{BF}^2  = -\frac{(d-1)^2}{4\,\ell_d^2}$ being the Breitenlohner-Freedman bound, one is allowed two possible choices \cite{Breitenlohner:1982jf, Klebanov:1999tb}. The scalar field can be taken to correspond to an operator of dimension either $\Delta_\pm$ in the hypothetical dual CFT$_{d-1}$. With the $\Delta_+$ ($\Delta_-$) choice, $A(x)$ ($B(x)$) is fixed as a boundary condition, and thus is treated as the source for the corresponding dual CFT operator. For $m_\text{eff}^2 > m_{BF}^2 +\ell_d^{-2}$ only the $\Delta_+$ mode is normalizable, and correspondingly one treats $A(x)$ as the source and $B(x)$ as the vacuum expectation value of the scalar operator in the dual theory.

\paragraph{Vectors ($s=1$):} The allowed boundary conditions for gauge fields are more intricate and depend on the dimension $d$. Of particular interest to us is the case $d =4$.  From  the analyses of \cite{Breitenlohner:1982jf,Witten:2003ya,Leigh:2003ez,Ishibashi:2004wx,Marolf:2006nd} it follows that one can impose at least two distinct boundary conditions for the gauge field. We will call these two boundary conditions Dirichlet/standard/electric or Neumann/modified/magnetic boundary conditions.  These two boundary conditions can be characterized in terms of the fall-off conditions of the gauge field. In the gauge $A_r =0$ in \AdS{4}, the Maxwell equations have as their general solution near the boundary
\begin{equation}
A_\mu (r,x) \xrightarrow{r\to \infty} a_\mu(x)  + \frac{b_\mu(x)}{r}  \ , \qquad x = \{ t, \Omega_2\},
\label{Afalloff}
\end{equation}
for any functions $a_{\mu}(x)$ and $b_{\mu}(x)$. The two cases are:	
\begin{itemize}
\item Dirichlet (standard): Here we fix the boundary value of the gauge field, i.e., we hold $a_\mu(x)$ fixed on \ESU{3}. With these boundary conditions one is allowed electrically charged states in the \AdS{d} geometry since flux lines can extend to the boundary, and constant gauge transformations act on the theory as a global symmetry. As is familiar from AdS/CFT, $a_\mu(x)$ acts as a source for the conserved current operator $J^\mu(x)$ of dimension $d-2$ on the \ESU{d-1}.
\item Neumann (modified): For these boundary conditions one holds $b_\mu(x)$ fixed, while allowing $a_\mu(x)$ to fluctuate. These boundary conditions can be realized by starting with the Dirichlet boundary conditions and integrating over all $a_\mu(x)$. With these boundary conditions one has a Gauss' law constraint in the theory; in fact, for $b_{\mu}=0$, the equation of motion of $a_\mu$ says that the associated current on \ESU{3} vanishes at every point, $\langle J^{\mu}(x)\rangle = 0$.
\end{itemize}
More general boundary conditions involving functional relations between $a_\mu(x)$ and $b_\mu(x)$ are also allowed; we refer the reader to \cite{Witten:2003ya,Marolf:2006nd} for a detailed account of these possibilities.

In all higher dimensions\footnote{For $d=3$, so long as the kinetic term is of pure Maxwell type, one must in fact fix $b_\mu$ on the boundary.  This is not familiar from AdS/CFT.  In the AdS/CFT context, the action for the AdS${}_3$ gauge fields generally contains Chern-Simons terms, which turn out to significantly affect the allowed boundary conditions.} $d > 4$, the only simple allowed boundary condition is to fix the analog of $a_\mu$ on the boundary.

\paragraph{Fermions ($s=\frac{1}{2}$):} For fermionic fields satisfying the Dirac equation with mass $m$, the boundary conditions again depend on $m$ \cite{Henningson:1998cd,Mueck:1998iz, Amsel:2008iz,Iqbal:2009fd}. A spinor in \AdS{d} corresponds to a fermionic operator of dimension $\Delta_\pm = \frac{d-1}{2}  \pm |m\,\ell_d|$ in the dual CFT. In the range $0 \le m\, \ell_d < \frac{1}{2}$ both of the $\Delta_\pm$ modes are normalizable, while for $m\,\ell_d \ge \frac{1}{2}$ only the $\Delta_+$ mode is normalizable.

Further details on how these boundary conditions are arrived at and what they imply for the spectrum of the theory on \AdS{d} can be found in \App{s:bcreview}.

\section{Four dimensional gauge theories on \AdS{4}}
\label{s:gaugeads4}

From the discussion of boundary conditions of free fields in \sec{s:qftads} it is clear that there is a wide variety of choices consistent with the $SO(d-1,2)$ isometry group of \AdS{d}. For our purposes the most interesting examples to consider are four dimensional gauge theories, which we now explore in some detail.

As we describe in \App{s:vectorbc}, from the analysis of \cite{Ishibashi:2004wx} it follows that the gauge-invariant information in the Maxwell equations can be summarized in terms of the dynamics of effective scalar fields. In \AdS{4} the spectral content of the theory is (almost) identical to that of two conformally coupled scalar fields; we used this to identify the possible boundary conditions for the gauge fields, which we called Dirichlet (standard) and Neumann (modified). In fact, for Maxwell electrodynamics on \AdS{4},  these boundary conditions are exchanged under the classical $SL(2,\ZZ)$ electric-magnetic duality of the bulk gauge theory
\cite{Witten:2003ya}.

To see the distinction between the two boundary conditions, recall that the standard boundary condition (Dirichlet) on a gauge field $A_\mu$ in \AdS{4} involves fixing its boundary value $A^\partial_{\mu} = a_\mu$ on the conformal boundary (the \ESU{3}). The path integral with these boundary conditions is interpreted in the AdS/CFT context as providing a generating functional for the global current correlators in the dual CFT$_3$. In other words the boundary values of the gauge field couple to the conserved current $J_\mu$ living on \ESU{3}, and one is computing  correlation functions in this hypothetical dual theory perturbed by a coupling $\int d^3 x \, a_{\mu} \,J^{\mu}$. The one-point function of this current is related to the sub-leading behaviour of $A^{\mu}$ near the boundary. Note that if one wants to switch off the sources for the currents, all one needs to do is demand that $a_\mu =0$, which can be phrased gauge-invariantly as demanding that the boundary components of the field strength vanish, $F_{\mu\nu}\mid_{\text{ESU}_3} = 0$.

With this choice the gauge field near the boundary is not a fluctuating degree of freedom, and in particular, as is clear from  above, one is  allowed to consider states charged under the gauge field -- there is no Gauss' law constraint arising from the boundary conditions. These charged states in the bulk correspond to states carrying a global symmetry charge in the putative dual CFT on \ESU{3}. These are familiar in the AdS/CFT context, and are frequently used in studies of charged black holes which model finite density states of the field theory.

The modified boundary condition (Neumann), on the other hand, fixes the sub-leading term in the expansion of $A_{\mu}$ near the boundary. This behaviour can also be described in terms of fixing the leading term in $F_{r \mu} $, but it does not fix the constant term of $A_\mu$ on the \ESU{3}. Since the constant term is not fixed, it is integrated over in the path integral.  We can just think of it as a boundary gauge field, i.e., we consider the dual CFT action perturbed as before by $\int d^{3}x\, a_{\mu} \,J^{\mu}$, and simply promote $a_\mu$ to be a field variable on \ESU{3}, to be integrated over in the path integral (as described initially in \cite{Witten:2003ya}).

However, this $a_\mu$ has no kinetic term.\footnote{Classically; such a kinetic term can generally be generated by quantum corrections.} Its promotion to a dynamical field in the boundary path integral  has the same effect as allowing any coefficient for the leading term of $A_{\mu}$, and effectively gauges the global symmetry. Of course, we still have the gauge symmetry acting on $A_{\mu}$ in the bulk, which also acts  on $a_{\mu}$ in the standard way required for a boundary gauge field. It may seem that in this description we did not put in the modified boundary conditions saying (in the simplest case) that the sub-leading term in $A_{\mu}$ should vanish; but in fact, now the equation of motion of $a_{\mu}$ makes $\vev{J_{\mu}}=0$ which is precisely the statement of the modified boundary condition.  In effect, these boundary conditions are entirely equivalent to gauging the boundary value of the gauge field. In particular, we have a Gauss' law in this case on the boundary, and the charge on the boundary is the same as the charge in the bulk (since the boundary gauge field is just the boundary limit of the bulk gauge field).

Our discussion thus far has been for Maxwell dynamics in \AdS{4}, but one can consider these general boundary conditions also for non-Abelian gauge fields with  gauge group $G$. Now we have a wide variety of choices for the boundary conditions to impose on the gauge field.  The two obvious ones are to impose Dirichlet or Neumann boundary conditions for all components of the gauge field. In the former case we can consider states charged under $G$ in the bulk \AdS{4}, while in the latter case one has a Gauss' law constraint forbidding states carrying $G$-charge. But we can also pick a sub-group  $H \subset G$ and impose Neumann boundary conditions for gauge fields in $H$, and Dirichlet for the others. We then have an $H$ gauge group in the full theory, and the Gauss' law forbids states charged under $H$ (but allows some states charged under $G$ which are neutral under $H$). We can think of this as gauging (in the three dimensional sense) a subgroup $H$ of the global symmetry group $G$.

In \cite{Witten:2003ya} the existence of these boundary conditions for a $U(1)$ gauge theory was argued based on the electric-magnetic S-duality symmetry. For arbitrary gauge group $G$ one would not have this symmetry unless matter fields of appropriate representation content were present. However, the discussion of \App{s:vectorbc} which summarizes the analysis of \cite{Ishibashi:2004wx,Marolf:2006nd} implies that one is free to choose these boundary conditions for any matter content. The crucial fact to remember is that dynamics in \AdS{4} is incomplete without specification of boundary conditions, due to the background being non-globally hyperbolic, and any set of boundary conditions that yields a well-defined phase space with a sensible Cauchy evolution should be admissible.

\section{$\CN =4$ SYM on \AdS{4} at weak coupling}
\label{s:n4ads}

Having discussed various aspects of boundary conditions for field theories on \AdS{d} spacetimes, we now turn to a specific example, the dynamics of $\CN=4$ SYM on \AdS{4} (both at zero temperature and at finite temperature). We first discuss some aspects of $\CN=4$ SYM on \AdS{4}  at weak coupling, and compute the free field partition function in this section. This can then be contrasted with the following sections where we will describe the physics at strong coupling. Everything we say in this section about the free theory can be easily generalized to any four dimensional gauge theory, but we discuss in detail the $\CN=4$ SYM example since
in that case we can also understand the theory at strong coupling.

\subsection{Spectrum of free $\CN =4$ SYM on \AdS{4}}
\label{s:freespec}

 The $\CN =4$ SYM theory has six scalar fields $\phi^I$ ($I=1,\cdots,6$) transforming in the adjoint representation of the gauge group $G$,  and four Weyl fermions $\psi^a$ ($a=1,2,3,4$). The Lagrangian is (our conventions for the $\sigma$ matrices are as in \cite{Beisert:2004ry}) :
\begin{eqnarray}
\CL_{\CN =4} \!\!\!&=& \frac{1}{g_{YM}^2} \, \Tr{\frac{1}{4} \, F_{\mu\nu}\, F^{\mu\nu}  + \frac{1}{2} D_\mu \phi^I \, D^\mu \phi^I + \sum_{I>J} \, [\phi^I , \phi^J]^2 + \frac{R}{12}\, (\phi^I)^2}\nonumber \\
 && \hspace{-5mm}+ \Tr{{\psi^{\dagger}}^a_{\dot \alpha}  \, \sigma^{{\dot \alpha}\beta}_\mu\, D^\mu \psi_{\beta a} - \frac{i}{2}
 \psi_{\alpha a} \, \sigma^{ab}_K \, \epsilon^{\alpha \beta}\, [\phi^K,\psi_{\beta\,b}]
 - \frac{i}{2}\, {\psi^{\dagger}}_{{\dot\alpha}}^a \, \sigma_{Kab} \, \epsilon^{{\dot\alpha}{\dot\beta}}\, [\phi^K,{\psi^\dagger}_{{\dot\beta}}^b]
 } \!\!.
\label{nfourlag}
\end{eqnarray}	
Note that we have accounted in the Lagrangian for the  conformal coupling of the scalars to the background curvature $R$. We will find it convenient to refer  to the collection of the basic fields in the theory by $\Psi = \{A_\mu, \phi^I, \psi_{\alpha a}\}$.

We are interested in knowing the spectrum of the free theory. The scalar fields satisfy a conformal Klein-Gordon  equation and so we have from \App{s:scalarbc}
\begin{equation}
\omega_{\phi^I}  = (\Delta_\pm + 2\,n + k) \,  \ell_4^{-1}\ , \qquad \Delta_\pm =1,2\quad \text{with}\;\; n,k\in \ZZ_+,
\label{n4scaspec}
\end{equation}	
where the choice of $\Delta_\pm$ is determined by our choice of normalizable mode, $k$ is the angular momentum quantum number, and states are $(2\,k+1)$-fold degenerate. The fermions in the theory are massless and therefore their spectrum is given by (independently of the boundary conditions)
\begin{equation}
\omega_{\psi^\alpha}  = (\Delta + 2\,n + k) \,  \ell_4^{-1}\ , \qquad \Delta =\frac{3}{2} \quad \text{with}\;\; n,k\in \ZZ_+.
\label{n4ferspec}
\end{equation}	
Again one has $(2\,k+1)$ states at a given angular momentum level $k$.

The spectrum of vector fields can easily be computed by realizing that the scalar ($A_\mu^{\bf s}$) and vector ($A_\mu^{\bf v}$) parts of $A_\mu$, satisfy the conformally coupled scalar wave equation \cite{Ishibashi:2004wx}. This implies that the spectrum is determined by the scalar spectrum in  \req{n4scaspec}. However, now we also have to account for the allowed boundary conditions. Using \App{s:bcreview} to translate the discussion of \sec{s:gaugeads4} one finds
\begin{itemize}
\item standard (Dirichlet):  Vector modes are treated as  $\Delta =2$ and the scalar modes are treated as
$\Delta =1$.
\item modified (Neumann): Vector modes are treated as  $\Delta =1$ and the scalar modes are treated as
$\Delta =2$.
\end{itemize}
Furthermore, it turns out that the vector fields lead only to the $k \ge 1$ part of (\ref{n4scaspec}), due to the well-known absence of spherically-symmetric radiation for vector gauge fields; see \cite{Ishibashi:2004wx} and \App{s:vectorbc} for details.

Note also that the degeneracy of the vector and scalar modes in \AdS{4} is the same since the vector and scalar harmonics on $\Sp^2$ are related via (see \App{s:bcreview} for the definitions of the harmonics)
\begin{equation}
{\bf V}^{(k)}_i  = \epsilon^{ij} \, \nabla_j \, {\bf Y}^{(k)} \ \ .
\label{}
\end{equation}	

 Generally one can choose to impose either Neumann ($\Delta_-$ fall-off) or Dirichlet ($\Delta_+$ fall-off) boundary conditions on the scalar fields in the $\CN=4$ supermultiplet. In the following we will usually implement this choice by requiring $0\le \alpha \le 6$ of the scalars $\phi^I$ to have Dirichlet boundary conditions and the remaining to satisfy Neumann boundary conditions.   However, a generic such choice will break all the supersymmetries of the theory. As was shown originally in \cite{Breitenlohner:1982jf}, supersymmetry requires that one has an equal number of $\Delta_+$ and $\Delta_-$ scalars, i.e., $\alpha =3$ in the notation introduced above.

\subsection{Free $\CN = 4$ SYM partition functions}
\label{s:freepartfn}

We have now assembled the data required to compute the spectral information  of $\CN =4$ SYM on \AdS{4}. We will proceed to compute the free field partition function directly for the various choices of boundary conditions described earlier.  It is first useful to record the partition function of the `letters', i.e., the basic fields $\Psi=\{A_\mu, \phi^I, \psi_{\alpha a}\}$ of the theory. We will refer to these, following the earlier analysis of gauge theories on compact spatial manifolds, as single particle partition functions \cite{Aharony:2003sx}. These will then be used to compute the free energy of the theory at finite temperature with various choices of boundary conditions.

\subsubsection{Single particle partition sum}
\label{s:singpart}

The spectral data for the free theory is sufficient to derive single particle partition sums, or equivalently the partition function of the Abelian theory with $G=U(1)$, for the free theory. We define this quantity as
\begin{equation}
z(x) = \sum_{\text{single-particle states}} \sum_\omega \, e^{-\beta\, \omega} \ , \qquad x \equiv e^{-\beta\, \ell_4^{-1}} \ .
\label{}
\end{equation}	

Consider first the single particle partition function for the adjoint-valued scalar fields of the theory. From the spectrum \req{n4scaspec}, we find
\begin{equation}
 z^\pm_{\text{scalar}}(x) = x^{\Delta_\pm} \, \sum_{n,k =0}^\infty \, (2 \, k +1) \, x^{2\, n+k}  = \frac{x^{\Delta_{\pm}}} {(1-x)^3} \ , \qquad \Delta_\pm = 1,2  \,,
 \label{sca1p}
\end{equation}	
where we have left the answer in terms of $\Delta_\pm$ to leave open the possibility of imposing different boundary conditions on each of the scalars. Likewise we can compute the single particle partition sum for the fermions. $\CN =4$ SYM has 4 Weyl fermions, each of which has 2 degrees of freedom. For a single Weyl fermion with spectrum \req{n4ferspec} one has
\begin{equation}
z_\text{fermions}(x)  =2\, x^{\frac{3}{2}} \, \sum_{n,k =0}^\infty \, (2 \, k +1) \, x^{2\, n+k}  = 2\, \frac{x^{\frac{3}{2}}}{(1-x)^3} \ .
\label{fer1p}
\end{equation}	

The single particle partition sum for the gauge fields is also easy to compute. However, we now have to account for the scalar ($A_\mu^{\bf s}$) and vector ($A_\mu^{\bf v}$) degrees of freedom and also for the boundary conditions discussed earlier. As discussed above and in \App{s:vectorbc}, for either Dirichlet or Neumann boundary conditions the gauge field spectrum is precisely the $k \ge 1$ part of the spectrum for a pair of scalars, one having $\Delta =1$ and the other having $\Delta =2$.  Thus we have\footnote{This answer can also be derived by counting the number of conserved current operators and their descendants in a $2+1$ dimensional CFT (the hypothetical dual living on \ESU{3}).}
\begin{equation}
z_{\text{gauge}}(x) = \sum_{k=1}^\infty \, \sum_{n=0}^\infty \,(2\,k+1)\, \left(x^{2} +x \right)\, x^{2\,n + k} =\frac{3\, x^2 - x^3}{(1-x)^3} \ \ \ .
\label{gauge1p}
\end{equation}	

Putting all this information together, we find that the single letter partition function for free $\CN =4$ SYM  (with boundary conditions parameterized by $\alpha$ as above) is given by
\begin{equation}
z(x) = \alpha\,  z^+_\text{scalar}(x) + (6-\alpha)\,z^-_\text{scalar}(x) + 4\,  z_\text{fermions}(x) + z_\text{gauge}(x).
\label{z1p}
\end{equation}	
%

\subsubsection{Multi-particle partition sum}
\label{s:multipart}

 In a relativistic CFT, the sensible partition sum to compute is for multi-particle states.  This can be easily derived by suitable combinations of the single particle partition sums. So, now we have to finally face the issue of how the boundary conditions influence this computation.

\paragraph{Multi-particle partition function for Dirichlet bc:} Let us first discuss the case where we impose Dirichlet boundary conditions on the gauge fields. There is no Gauss' law constraint for electrically charged states in the theory, and one can therefore simply compute the multi-particle partition function by summing over multi-particle states, keeping track of indistinguishability of the particles and accounting for the statistics.

For each bosonic degree of freedom, the multi-particle partition sum is given by
\begin{equation}
Z_\text{boson}(\beta) = \prod_{\omega} \, \left(\frac{1}{1-e^{-\beta\, \omega}}\right)\,,
\label{bosmpart}
\end{equation}	
where $\omega$ are the single particle energy levels. Using the spectrum of the single particle states in \AdS{4} \req{n4scaspec}, one finds for each component of the adjoint valued scalars in $\CN =4$ the partition sum:
\begin{eqnarray}
Z^\pm_\text{scalar} (x) &=& \prod_{n,k=0}^{\infty} \, \left(\frac{1}{1-x^{\Delta_\pm + 2 \, n + k}}\right)^{2k+1} = \exp\left(\sum_{p=1}^\infty \, \frac{1}{p}\,  z^\pm_\text{scalar}(x^p)\right) \nonumber \\
&=& \prod_{m=0}^\infty \, \left(\frac{1}{1-x^{\Delta_\pm+m}}\right)^{\frac{1}{2}\,(m+1)\,(m+2)} \ \ \ .
\label{mzscalar}
\end{eqnarray}	
The partition sum for the vectors can be obtained similarly, using the
single-particle partition function \req{gauge1p}; one has
\begin{equation}
Z_\text{gauge} (x) =\exp\left(\sum_{p=1}^\infty \, \frac{1}{p}\,  z_\text{gauge}(x^p)\right) =\prod_{m=0}^\infty \,
\left(\frac{1-x^{m+3}}{(1-x^{m+2})^3}\right)^{\frac{1}{2}\,(m+1)\,(m+2)}.
\label{mzvector}
\end{equation}	
Fermions on the other hand have a multi-particle partition sum which reads
\begin{equation}
Z_\text{fermion}(\beta) = \prod_{\omega} \, \left(1+e^{-\beta\, \omega}\right).
\label{fermpart}
\end{equation}	
From the spectrum \req{n4ferspec} one derives
\begin{eqnarray}
Z_\text{fermion} (x) &=& \prod_{n,k=0}^{\infty} \, \left(1+x^{\frac{3}{2}+ 2 \, n + k}\right)^{2\, (2k+1)} = \exp\left(\sum_{p=1}^\infty \, \frac{(-1)^{p+1}}{p}\,  z_\text{fermions}(x^p)\right) \nonumber \\
&=& \prod_{m=0}^\infty \, \left(1+x^{\frac{3}{2}+m}\right)^{(m+1)\,(m+2)} \ \ \ .
\label{mzfermion}
\end{eqnarray}	

Putting these pieces together one finds that the partition function for $\CN =4$ SYM in \AdS{4} is given by
\begin{equation}
Z_{\CN =4} (x)= \bigg[\left(Z^+_\text{scalar}(x)\right)^\alpha \, \left(Z^-_\text{scalar}(x)\right)^{6-\alpha} \, Z_\text{gauge}(x)\,Z_\text{fermion}^4(x)\bigg]^{\text{dim}({\mathfrak g})},
\label{muldir}
\end{equation}	
where we have accounted for the fact that there are $\text{dim}({\mathfrak g})$ degrees of freedom in the adjoint valued fields in the gauge group $G$. Since each of these states carrying the gauge charge can be treated separately, we have simply exponentiated the final result of the Abelian theory.

The asymptotic behaviour of $\log(Z_{\CN =4} (x))$ for high temperatures $x \to1$ can be extracted by saddle point methods.\footnote{A suitable generalization of a theorem due to Meinardus allows determination of the asymptotics of $Z_{\CN =4}$ itself, see \cite{Lucietti:2008cv}.} In particular, the free energy of the theory defined as usual as
\begin{equation}
F = -T \, \log (Z_{\CN =4} (\beta))
\label{}
\end{equation}	
behaves for $T\, \ell_4 \gg 1$ as
\begin{equation}
F = -15\, \zeta(4)\, T^4\, \ell_4^3\, \text{dim}({\mathfrak g}) = -\frac{\pi^4}{6}\, T^4\, \ell_4^3\, \text{dim}({\mathfrak g}) \ .
\label{Fdir}
\end{equation}	

The fact that the free energy \req{Fdir} shows a characteristic $T^4$ behavior at high temperatures, $\beta\, / \ell_4 \ll 1$, is in keeping with the physical idea that the asymptotic part of the spectrum should be insensitive to the \AdS{} curvature. However, due to the confining nature of the \AdS{} spacetime, the free energy is exponentially damped at temperatures $T\, \ell_4 \ll 1  $. Nevertheless, note that the free energy obtained from \req{muldir} scales with $\text{dim}({\mathfrak g})$ (which determines the central charge) for all temperatures. This scaling of the free energy is easy to understand, once one notes the absence of the Gauss' law constraint in \AdS{4} for electrically charged states.  Since we are allowed states carrying arbitrary $G$-valued charges, we should allow them in our partition sum. As a result each adjoint valued field in the theory acts as a distinguishable particle and the free energy simply scales as the number of such fields.

\paragraph{Multi-particle partition function for Neumann bc:} With the Neumann boundary conditions one has to deal with the Gauss' law constraint for the $G$-valued gauge field. As described in \sec{s:gaugeads4}, one way to impose these boundary conditions is to gauge the boundary value of the bulk $A_\mu$. One then has a $G$-valued boundary gauge field living on \ESU{3} leading to Gauss' law, which in particular implies that we only have singlet states in the bulk.

The free gauge theory partition function with the Neumann boundary condition can be computed using the methods described in \cite{Sundborg:1999ue,Polyakov:2001af, Aharony:2003sx}; we will follow the treatment of \cite{Aharony:2003sx} in what follows. To enumerate the gauge-invariant operators in the theory, we should construct gauge-invariant words from the basic letters. So we string along the fields from the set $\Psi$, each weighed by its energy, and project onto singlet states to achieve gauge invariance. The projection has to be done accounting for particle statistics; as usual we pick symmetric combinations of the bosons $\Psi_B = \{ A_\mu,\phi^I\}$ and anti-symmetric combination of the fermions $\Psi_F = \{\psi_{\alpha a}\}$. As explained in \cite{Aharony:2003sx} we can express the result in terms of an integral over the gauge group. One has
\begin{equation}
Z(x) = \int[DU] \, \exp\bigg(\sum_{m=1}^\infty  \, \frac{1}{m}\, \left[z_B(x^m)  + (-1)^{m+1} \, z_F(x^m)\right] \, \chi^G_\text{adj}(U^m) \bigg),
\label{Zneu}
\end{equation}	
where $U$ is a group element of the gauge group $G$ and $[dU]$ is the standard Haar measure on the group manifold. $\chi^G_\text{adj}$ is the group character in the adjoint representation (since all the field content of $\CN =4$ SYM transforms in the adjoint of  $G$).

In order to account for the statistics, we have split up the single particle (or Abelian) partition function into bosonic and fermionic parts. To be precise,
\begin{equation}
z_B(x) = \alpha\,  z^+_\text{scalar}(x) + (6-\alpha)\,z^-_\text{scalar}(x) + z_\text{gauge}(x) \ , \qquad z_F(x) =  4\,  z_\text{fermions}(x)\,.
\label{zbzf}
\end{equation}	

 Note that, like other theories with a discrete spectrum and a Gauss' law constraint, the theory exhibits a large $N$ phase transition between a ``confining'' phase with a free-energy of order one, and a ``deconfined phase'' which behaves like a free theory in the bulk, i.e., $F \sim \ord{N^2}$ \cite{Sundborg:1999ue,Aharony:2003sx}.  This transition happens precisely at the temperature \footnote{At large $N$ the computation can be carried out more simply by enumerating the gauge-invariant operators. The transition temperature is determined by examining the limit of convergence of the partition sum thus computed. Equivalently one can re-express \req{Zneu} in terms of the eigenvalue distribution of $U$ and look for the boundary of stability of the uniform distribution (see also \App{s:NeuHag}).}
\begin{equation}
T_\star = -\frac{1}{\ell_4\, \log (x_\star)} \  ,\; \text{with} \qquad z(x_\star) = z_B(x_\star) + z_F(x_\star) = 1\,.
\label{weakHagA}
\end{equation}	
The low temperature phase has a Hagedorn density of states proportional to $\exp(E/T_\star)$.  One can evaluate the location of this transition for various choices of scalar boundary condition specified by $\alpha$. The result is tabulated in Table \ref{t:xstardat}. At weak non-zero 't Hooft coupling $\lambda$, the transition either remains a first
order phase transition or splits into two continuous phase transitions; a three-loop computation
is required to distinguish between these two possibilities \cite{Aharony:2003sx}.
\begin{table}[htdp]
\begin{center}
\begin{tabular}{|c|c|}
\hline
$\alpha$ & $x_\star$ \\
\hline
0 & 0.144692  \\
1 & 0.132126 \\
2 & 0.120972 \\
3 & 0.111111 \\
4 & 0.102408 \\
5 & 0.0947246\\
6 & 0.0879331 \\
\hline
\end{tabular}
\caption{The location of the Hagedorn transition in large $N$, $SU(N)$ $\CN =4$ SYM on \AdS{4} as a function of $\alpha$, which determines the boundary condition on the scalar fields in the theory.}
\label{t:xstardat}
\end{center}
\end{table}

Note that the symmetric choice of scalar boundary conditions $\alpha =3$, which can preserve supersymmetry  (in the ground state), leads to a particularly simple value for the transition temperature, $T_\star= 1 / (2\, \log(3) \,\ell_4)$.

\paragraph{Multi-particle partition function for Neumann boundary condition for $H \subset G$:}
One can also perform a similar analysis when we have modified boundary conditions for a subgroup $H$ of $G$. Now one has a singlet constraint on $H$-valued fields. Assuming that the fields in the complement $G/H$ transform in representations $R_i$ of $H$ one obtains the partition function:
\begin{equation}
Z(x)
= \int[DU] \, \exp\bigg(\sum_{m=1}^\infty  \, \frac{1}{m}\, \left[z_B(x^m)  + (-1)^{m+1} \, z_F(x^m)\right] \, \left( \chi^H_\text{adj}(U^m) + \sum_i \chi^H_{R_i}(U^m) \right) \bigg),
\label{HNeupf}
\end{equation}	
where now $U \in H$. Note that those fields in the complement $G/H$ which transform trivially under $H$ will essentially have Dirichlet boundary conditions. On general grounds one expects that the result for the partition sum then simply reduces to the multi-particle partition sum \req{muldir} calculated earlier, which one can check it indeed does (since $\chi=1$ in that case).
Depending on the precise representation content under $H$, when this group has a rank of order $N$, we may again obtain a large $N$ phase transition similar to the one described above. This will be analyzed for some specific cases in \S\ref{s:involweak} below.

\section{$\CN=4$ SYM on AdS$_4$ at strong coupling}
\label{s:strongN4}

We have thus far concentrated on the dynamics of free $\CN =4$ SYM theory on \AdS{4}, and seen that the physics is very sensitive to the choice of boundary conditions. Since the theory has a marginal coupling constant $g_{YM}^2$, it is interesting to examine the behaviour when one is far removed from the free theory. There are two reasons that this is a tractable problem. One is that the field theory has a quantum $SL(2,\ZZ)$ electric-magnetic duality, which allows one to map the strongly coupled electric description to a weakly coupled magnetic description. For the theory defined as usual on  Minkowski spacetime, $ {\bf R}^{3,1}$, the S-duality maps the complexified coupling constant $\tau = \frac{\theta}{2\pi} + i\, \frac{4\pi}{g_{YM}^2}$ via $\tau \to -\frac{1}{\tau}$. This permits one to access the strongly coupled limit $g_{YM} \gg 1$ of the theory; for the present case we need to also understand the map of the boundary conditions under the S-duality action.

Another reason for considering the $ {\cal N} = 4$ SYM theory at strong coupling is that the theory on Minkowski spacetime or on \ESU{4} is known to have a holographic dual in terms of type IIB string theory on \AdS{5} $\times \ {\bf S}^5$ (in Poincar\'e and global coordinates, respectively), which is tractable in the large $N$ limit with large 't Hooft coupling \cite{Maldacena:1997re}.  But the AdS/CFT correspondence is general and allows one to consider the theory on other backgrounds as well (see below).
We would like to use
these two approaches to accessing the strong coupling regime of $ {\cal N}=4$ SYM on \AdS{4} to gain insight into the dynamics. We  will next take a preliminary stock of what one can hope to learn about the strongly coupled theory, before proceeding to discuss the details in the subsequent sections.

\paragraph{S-duality:} In order to understand the implications of S-duality, we need to know the action of the S-transformation of $SL(2,{\mathbb Z})$ on the boundary conditions. As we have already mentioned, there is a certain tension in extrapolating the result for the Abelian theory, where the Dirichlet and Neumann boundary  conditions are exchanged under S-duality, to the non-Abelian case. In particular, since in the non-Abelian case the two boundary conditions lead to a different global symmetry group, they can no longer be related. To get a precise statement, we need to understand the boundary conditions better, and fortunately this is possible for the class that preserve some amount of supersymmetry.  We will be led  towards the rich class of half-BPS boundary conditions which were recently discussed for $\CN =4$ SYM on the half-space by Gaiotto and Witten \cite{Gaiotto:2008sa}. These boundary conditions can in fact be ported immediately to the \AdS{4} geometry (which is conformally related); we will describe them and their implications in \sec{s:bc2}.

\paragraph{Holography and strong coupling:} Recall that  the AdS/CFT correspondence posits a duality between $\CN=4$ SYM on a background $\CB_4$ and Type IIB string theory on a ten dimensional spacetime ${\cal M}_{10}$ that asymptotes (locally)  to \AdS{5} $\times \ {\bf S}^5$, with a boundary conformally related to $\CB_4$.\footnote{This is the first occurrence in this paper of an asymptotically AdS spacetime in which gravity is dynamical. We will try to avoid confusion between this negatively curved geometry and the non-dynamical AdS background on which the field theory lives, by referring to the latter as the boundary manifold and the former as the bulk manifold.} The familiar examples of the AdS/CFT correspondence involve situations where ${\cal M}_{10}$ is a direct product of a negatively curved five manifold times a five sphere, but generically one will not have this factorized structure.

We are interested in the situation where ${\cal B}_4$ is itself a negatively curved geometry, ${\cal B}_4 =$ \AdS{4}. The \AdS{5} part of the geometry attained asymptotically should have as its boundary the \AdS{4} geometry we have chosen, with its prescribed metric $\gamma_{\mu\nu}$, up to a conformal factor. Denoting the characteristic scale of curvature of the bulk \AdS{5} geometry by $L_5$ (which is also the radius of curvature of ${\bf S}^5$, by the Type IIB equations of motion), one has the standard relation to the parameters of $\CN=4$ SYM with $G=SU(N)$ (up to numerical factors):
\begin{equation}
\lambda \equiv g_{YM}^2\, N = \frac{L_5^4}{\ell_s^4} \ , \qquad g_{YM}^2 = g_s\, ,
\label{holmap}
\end{equation}	
where we have defined the 't Hooft coupling $\lambda$ for the field theory. The correspondence thus allows one to access the regime of strong 't Hooft coupling in planar $\CN=4$ SYM, i.e., $\lambda \gg 1$ and $N \to \infty$. Finally, note that the relation between parameters given in \req{holmap} implies that the ten dimensional gravitational coupling $G_N^{(10)}$ is related to the central charge $c$ of the field theory; specifically
\begin{equation}
\frac{L_5^8}{16\pi\, G_N^{(10)} } \propto  c \ , \qquad \text{with}\;\; c \propto \text{dim}({\mathfrak g}) \ .
\label{GNc}
\end{equation}	

The question we face is what choice  of boundary conditions for $\CN=4$ SYM on \AdS{4} leads to a controlled holographic dual.  To work at the level of classical supergravity, one requires that the solution ${\cal M}_{10}$ to the Type IIB equations of motion be a smooth manifold. A necessary condition for this is that the number of light degrees of freedom\footnote{To be specific the constraint is that one not have a large number of low dimension gauge-invariant operators (say of ${\cal O}(N^2)$, as we have for the global symmetry currents in the case of Dirichlet boundary conditions) which can be activated by turning on appropriate sources. This is different from asymptotically AdS black hole geometries, which have  ${\cal O}(N^2)$ free energy, in that we are counting fields in the bulk, rather than states
(see \sec{s:holimp} for more details).} in the theory at large $N$ and strong 't Hooft coupling be $\CO(1)$, but this may not be sufficient.

\begin{figure}[htbp]
\begin{center}
\includegraphics[scale=0.75]{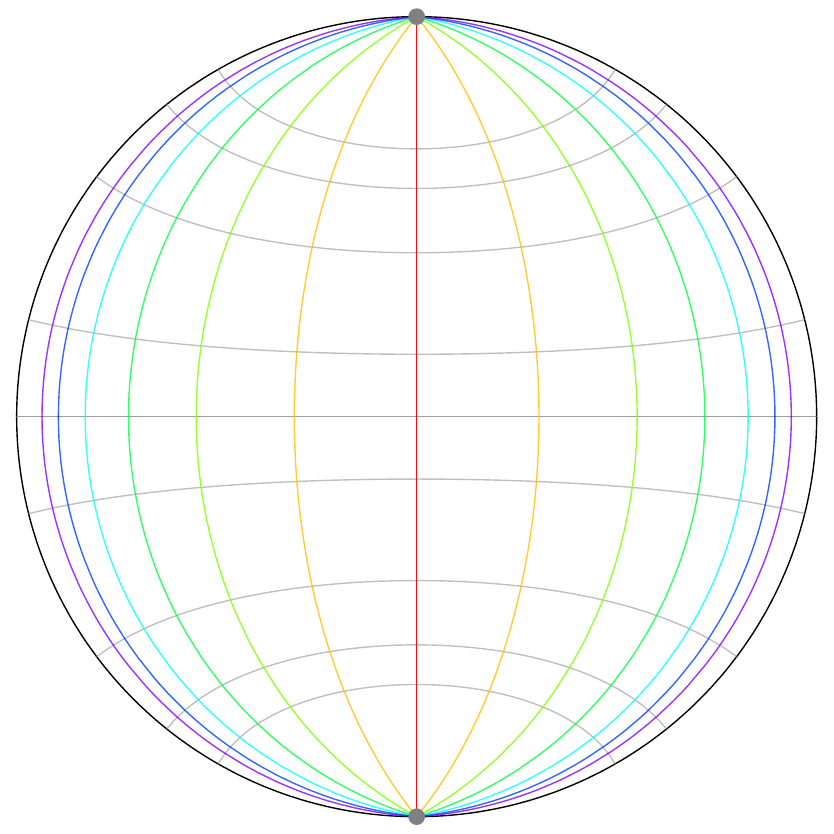}
\caption{Poincar\'e disk representation of \AdS{d} slices of \AdS{d+1} in the $(R,r)$ coordinates used in \req{adsinads}. The vertical curves (coloured) denote surfaces of constant $R$ which are \AdS{d} geometries, with the color coding showing the UV (boundary) and IR (interior).  The boundary of the spacetime is the edge of the disk, which is attained as $R\to \pm \infty$, and comprises of two copies of \AdS{d} joined together at their respective boundaries ($r\to \infty$). The horizontal curves (gray) are constant $r$ surfaces.}
\begin{picture}(0,0)
\setlength{\unitlength}{1cm}
\put(1.2,7.8){$R \to \infty$}
\put(1.2,7.58){\vector(1,0){1.5}}
\put(-2.5,7.75){$R \to -\infty$}
\put(-0.9,7.58){\vector(-1,0){1.6}}
\put(-0.42,10.75){$r=\infty$}
\put(-0.42,4.2){$r=\infty$}
\end{picture}
\label{f:adsinads}
\end{center}
\end{figure}
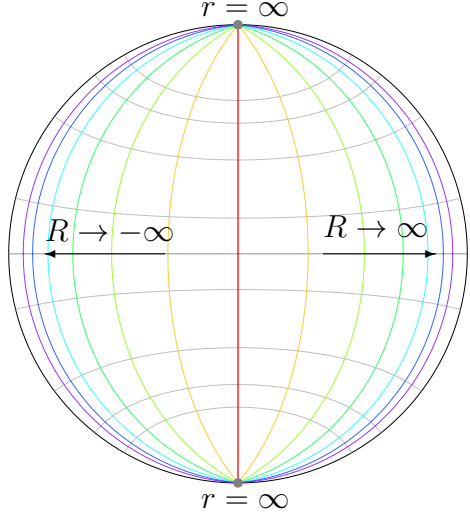

 A simple starting point which hints at an appropriate bulk spacetime is to consider the \AdS{5} $\times \ {\bf S}^5$ geometry, with the \AdS{5} part foliated by \AdS{4} slices, i.e., we pick the metric (suppressing for brevity the ${\bf S}^5$ part):
\begin{equation}
ds^2 = dR^2 + \frac{L_5^2}{\ell_4^2} \, \cosh^2\left(\frac{R}{L_5}\right)\, \gamma_{\mu \nu} \, dx^\mu\, dx^\nu  = \frac{L_5^2}{\cos^2(\Theta)} \, \left(d\Theta^2 + \frac{1}{\ell_4^2} \, \gamma_{\mu\nu}\, dx^\mu\,dx^\nu \right),
\label{adsinads}
\end{equation}
with $\gamma_{\mu\nu}$ being the standard metric on global \AdS{4} as given, for instance, in \req{global2}. However, it is important to note that in these coordinates the boundary is not a single copy of \AdS{4} but rather two copies of the same, attained as $R \to \pm \infty$ in the coordinatization chosen above. To see this explicitly, note that the global \AdS{5} geometry has the metric
\begin{equation}
ds^2 = -\left(1+\frac{\rho^2}{L_5^2} \right)\, dt^2 + \frac{d\rho^2}{1+\frac{\rho^2}{L_5^2}} + \rho^2\, (d\zeta^2 + \sin^2(\zeta)\, d\Omega_2^2),
\label{global5}
\end{equation}	
and one obtains \req{adsinads} by the coordinate transformation:
\begin{equation}
\frac{\rho^2}{L_5^2}  = \frac{L_5^2}{\ell_4^2} \, \cosh^2\left(\frac{R}{L_5}\right) \, \left(1+\frac{r^2}{\ell_4^2} \right)  -1\ , \qquad
\rho^2 \,\sin^2 (\zeta) = \frac{L_5^2}{\ell_4^2} \, r^2 \, \cosh^2\left(\frac{R}{L_5}\right).
\label{}
\end{equation}	
The coordinate ranges are $R \in (-\infty, \infty)$, $\Theta\in (-\pi/2,\pi/2)$
for \req{adsinads}, while $\rho \in [0,\infty)$ and $\zeta \in [0,\pi]$.
This coordinate transformation makes it clear that the \ESU{4} boundary of the global \AdS{5} is a double-cover of \AdS{4}. For $R>0$, the angular coordinate $\zeta$ runs from $0$ to $\pi/2$ while as $R$ goes negative, it explores the region $\pi/2$ to $\pi$.  Thus the two copies of \AdS{4} are joined across the equator of the ${\bf S}^3$, i.e.,  at  $\zeta = \pi/2$. See  \fig{f:adsinads} for illustration.

In \req{adsinads} we have an in principle candidate for the holographic dual of $\CN=4$ SYM on \AdS{4}, but it has two copies of \AdS{4} on its boundary instead of one.
One situation in which the geometry \req{adsinads} turns out to be appropriate is if one wishes to impose the so called transparent boundary conditions on the $\CN=4$ SYM fields. Implicitly, one takes the theory to be defined on the double-cover of global \AdS{4}, the \ESU{4}, for which it is well known that the holographic dual is global \AdS{5}, i.e., the geometry \req{global2} (with $d=5$). This was indeed the philosophy adopted in \cite{Hubeny:2009rc}.\footnote{The primary focus of that work was to understand the behaviour of strongly coupled CFTs on AdS black hole backgrounds. Here we are concerned with the dynamics in pure AdS spacetime, and in particular with the constraints imposed by the choice of boundary conditions.}

 For our current purposes, we wish to understand the behaviour of $\CN=4$ SYM on a single \AdS{4}.  As we will see below, if we want to describe the duals of the boundary conditions which preserve supersymmetry we also need to break the R-symmetry (at least down to $SO(3) \times SO(3)$).  In terms of supergravity solutions, this requires that we consider a more general ansatz for type IIB solutions, not simply a direct product spacetime  \AdS{5} $\times\ {\bf S}^5$ with a change of conformal frame at infinity as described above. In order to understand these issues better we discuss in the next section the possible choices of boundary conditions for $\CN=4$ SYM on \AdS{4}. We will see in the following sections that some
boundary conditions are indeed related to \req{adsinads}, after we identify the two copies of \AdS{4} at its boundary, but general boundary conditions require completely different holographic duals.

\section{Supersymmetric boundary conditions  for $ {\cal N} =4 $ SYM on \AdS{4}}
\label{s:bc2}

In our discussion of boundary conditions for $ {\cal N} =4$ SYM on \AdS{4} we have so far not used supersymmetry. One can easily engineer situations where the supersymmetry is broken via the boundary conditions. While these are interesting in their own right (the challenge being to engineer stable theories), we will focus here on examples of boundary conditions that preserve supersymmetry, in order to have tractable models, which one can analyze both at weak and strong coupling.

The first statement that should be made in this context is that the supersymmetry preservation necessitates an asymmetric treatment of the scalars $\phi^I$. It was pointed out in \cite{Breitenlohner:1982jf} that one should treat half of the scalars as corresponding to operators of dimension $\Delta = 2$, while the remainder have the $\Delta = 1$ boundary condition. This choice breaks the $SO(6)$ R-symmetry of the theory down to (at most) $SO(3) \times SO(3)$.\footnote{As described in \cite{Gaiotto:2008sa}, this follows algebraically by noting that the \AdS{4} superalgebra $OSp(4|4)$ has a $SO(4)  \simeq SU(2) \times SU(2)$ R-symmetry group.} The essential point is that supersymmetry transformations of a Dirac spinor relate it to both the $\Delta =1$ and $\Delta =2$ scalar representations.

A heuristic argument to intuit this choice is to note that, as described e.g., in \cite{Amsel:2008iz}, supersymmetric boundary conditions on \AdS{d} can be classified by organizing the boundary data into ${\cal N} = 1$ superfields on \ESU{d-1}.  Scalar superfields on \ESU{3} are not chiral, and both the constant and $\bar \theta \theta$ terms yield independent bosonic component fields.  Because the fermionic coordinate has dimension $-\frac{1}{2}$, the dimensions of these two bosons must differ by $1$.  As a result, preserving any amount of supersymmetry requires the bulk scalars to come in pairs with one member having $\Delta =1$ and the other having $\Delta =2$.

We will be interested here in boundary conditions that preserve sixteen supercharges, sitting in
the $OSp(4|4)$ algebra which is equivalent to the $d=3$, ${\cal N}=4$ superconformal algebra.
There are two ways to think about boundary conditions which preserve supersymmetry. From the \AdS{4} point of view, the first thing we need to do is to choose a supersymmetric vacuum for the ${\cal N}=4$ SYM on \AdS{4}. The theory in flat space has a moduli space of vacua, but this is lifted by the
conformal coupling of the scalar fields in (\ref{nfourlag}), so one may think that only the vacuum
at $\phi^I=0$ remains. But in fact there are many additional supersymmetric vacua, following from the
fact that the F-term condition for preserving supersymmetry in an ${\cal N}=1$ supersymmetric field theory on \AdS{4} is modified from $\partial W / \partial \Phi^i = 0$ (where $W$ is the superpotential and $\Phi^i$ are the chiral superfields) to
\begin{equation}
\frac{\partial W }{ \partial \Phi^i} + \frac{1}{ \ell_4}\, \frac{ \partial K }{ \partial \Phi^i} = 0 \ ,
\label{ssads}
\end{equation}	
 where $K$ is the K\"ahler potential. For the ${\cal N}=4$ SYM theory these equations take the form
\begin{equation}
[\Phi^i, \Phi^j] + \frac{1}{\ell_4}\,  \epsilon^{ijk}\, (\Phi^k)^{\dagger} = 0 \ ,
\label{adssy}
\end{equation}	
and these equations (together with the D-term equations) have many non-trivial solutions. For example, one can choose the real part of $\Phi^{1,2}$ and the imaginary part of $\Phi^3$ to be any $N$-dimensional representation of $SU(2)$ (multiplied by $1/\ell_4$).\footnote{It is amusing to note that these vacua are precisely the same as the classical vacua of the mass-deformed ${\cal N}=4$ SYM theory (also known as the ${\cal N}=1^*$ theory) in flat space \cite{Vafa:1994tf,Polchinski:2000uf}, with the role of the mass played by the AdS curvature. As we will see below, also the S-duality transformations of the vacua will bear a strong resemblance to that case. However, there are also many differences, and in particular our vacua preserve more supercharges that sit in a different algebra.} In such vacua the gauge symmetry is partly or completely broken, with some gauge fields becoming massive. Our analysis up to now assumed that we were expanding around the trivial vacuum, but the same arguments can be easily applied to any other supersymmetric vacuum; of course details of computations
like the partition functions of \sec{s:n4ads} will be modified. While the argument above only
shows that these vacua preserve ${\cal N}=1$ supersymmetry, the vacua corresponding to $SU(2)$ representations as described above in fact preserve all sixteen
supercharges. We will not analyze this in detail here, since we will momentarily show that an equivalent analysis has already been performed, so that we can just use the known results. In any case, after we
choose a supersymmetric vacuum, we still have generically some unbroken gauge symmetry, and we can then
choose Neumann boundary conditions for the gauge fields in some subgroup of the unbroken gauge symmetry (and Dirichlet boundary conditions for the other massless gauge fields; massive gauge fields necessarily have Dirichlet boundary conditions). We can also couple the
${\cal N}=4$ SYM to any degrees of freedom on the boundary (possibly charged under the unbroken
gauge group), as long as they preserve supersymmetry.

An alternative way to find all
supersymmetric boundary conditions on \AdS{4} (preserving sixteen supercharges) is to look at the allowed boundary conditions on the half-space, as done by Gaiotto and Witten \cite{Gaiotto:2008sa}.  As discussed there, there are a host of boundary conditions that preserve 8 ordinary and 8 superconformal supersymmetries of ${\cal N}=4 $ SYM. These boundary conditions can be naturally ported over to \AdS{4}, since this space is related by a conformal transformation to the half-space.
To see this, let us recall that the half-space $ {\bf R}^{2,1} \times {\bf R}_+ $ parameterized by $\{\tau, x^1, x^2, z\}$ with $z \ge 0$ has the canonical flat metric:
\begin{equation}
ds^2 = -d\tau^2  + (dx^1)^2 + (dx^2)^2 + dz^2.
\label{}
\end{equation}	
 This can be mapped into the metric of \AdS{4} in Poincar\'e coordinates by the conformal factor $\ell_4^2/z^2$, i.e.,
\begin{equation}
ds^2 = \frac{\ell_4^2}{z^2} \left(-d\tau^2  + (dx^1)^2 + (dx^2)^2+ dz^2\right).
\label{poinmet}
\end{equation}	
Passage to global \AdS{4} is attained by the map (obtained easily by the embedding of \AdS{4} in $ {\bf R}^{3,2}$):
\begin{eqnarray}
z =\frac{\ell_4}{\sqrt{\frac{r^2}{\ell_4^2} +1} \, \cos (\frac{t}{\ell_4})+ \frac{r}{\ell_4}\, \cos (\theta)} \ , &&\quad
\tau= \frac{\ell_4\, \sqrt{\frac{r^2}{\ell_4^2}  +1} \, \sin (\frac{t}{\ell_4}) }{\sqrt{\frac{r^2}{\ell_4^2}  +1} \, \cos (\frac{t}{\ell_4}) + \frac{r}{\ell_4}\, \cos (\theta)} \ ,
\nonumber \\
x^1 + i \, x^2 &=& \frac{ r \, \sin (\theta)\, e^{i\phi} }{\sqrt{\frac{r^2}{\ell_4^2} +1} \, \cos (\frac{t}{\ell_4})+ \frac{r}{\ell_4}\, \cos (\theta)} \ .
\label{gtopA}
\end{eqnarray}	
The transformation \req{gtopA} maps \req{poinmet} into the global \AdS{4} metric  \req{global2} (though it only covers a finite range of values of $t$). Putting the pieces together we see that, up to the conformal factor $z^2$ (which can be expressed in terms of the global coordinates using \req{gtopA}), one can indeed conformally map the half-space $ {\bf R}^{2,1} \times {\bf R}_+ $ into \AdS{4}. The boundary of the half-space $z = 0$ maps to the boundary of \AdS{4} located at $r \to \infty$; we thereby have a clear way to map the boundary conditions on the half-space into those for \AdS{4}.

Under the conformal mapping, the supersymmetries preserved  by the boundary conditions for fields on the half-space carry over into the supersymmetries of \AdS{4}.  Since the conformal group on the half-space (preserving its boundary) is simply $SO(3,2)$, which is the isometry algebra of \AdS{4}, both Killing and conformal Killing vectors (and  likewise spinors) on the half-space map to just Killing vectors (equivalently spinors) on \AdS{4}.  Since the Gaiotto-Witten boundary conditions we describe preserve $8$-supercharges and $8$-superconformal charges on the half-space, by mapping them over to the \AdS{4} geometry we obtain boundary conditions preserving 16 regular supercharges.

We will now review the boundary conditions which preserve supersymmetry on the half-space $ {\bf R}^{2,1} \times {\bf R}_+ $ as this will prove useful to discuss the strong coupling behaviour. Our notation will for the most part follow that of \cite{Gaiotto:2008sa}, although we  make a few minor notational changes (such as the half-space  coordinate being $z$ with the boundary at $z =0$). Since the simplest SUSY preserving boundary conditions treat the adjoint scalars of $\CN=4$ SYM asymmetrically, we will denote the two sets of scalars as $X^i$ and $Y^i$ ($i=1,2,3$), following \cite{Gaiotto:2008sa}. We will first present the most general result for the supersymmetric boundary conditions for ${\cal N} =4$ SYM on the half-space, and then move to examples.

The most general $\frac{1}{2}$-BPS boundary condition for ${\cal N} =4$ SYM on the half-space $ {\bf R}^{2,1} \times {\bf R}_+$ can be described \cite{Gaiotto:2008sa} in terms of a triple $ \left(\rho, H, {\mathfrak B} \right)$. This data is to be understood as follows: start with a gauge group $G$ and first provide a homomorphism $\rho$  from $SU(2)$ into  ${\mathfrak g}$. The map $\rho$ breaks the gauge group $G$ down to $H$, which is the part of the gauge symmetry that is preserved by the boundary conditions. One therefore has $H$-valued gauge fields which live on the boundary of the half-space, which can be coupled to a $2+1$ dimensional theory ${\mathfrak B}$ having $H$ as its global symmetry group.  We will call ${\mathfrak B}$ the boundary CFT in what follows. We will see in \sec{s:bcquotients} below that this data can be refined a bit further; the gauge symmetry $G$ may be broken by a choice of involution in addition to that caused by choice of  homomorphism $\rho$.

We will see in the next subsection how the choice of $\rho$ is related to the choice of supersymmetric
vacuum on \AdS{4} mentioned above. Examples of supersymmetric boundary conditions can be easily described in terms of brane constructions, when the gauge group $G$ is a unitary group $U(N)$ for some $N$. In such cases we can think of the ${\cal N} =4$ SYM arising on the world-volume of $N$ D3-branes oriented along (say) $0123$ in ${\bf R}^{9,1}$. To implement the half-space boundary conditions we will  use D5-branes (oriented along $012456$) and NS5-branes (oriented along $012789$); the  D3-branes end on these five-branes and we will see all three pieces of data described above very cleanly in this picture.

To keep things simple let us first start with the most basic boundary conditions where ${\mathfrak B}$ is trivial, i.e., we do not have explicit boundary degrees of freedom. There are two basic choices of half-SUSY boundary conditions that one can impose:

\begin{itemize}
\item Dirichlet or D-type: where we impose Dirichlet boundary conditions on the gauge fields and one half of the scalars (say $Y$), i.e., $F_{\mu\nu}$ and $Y$ vanish on the boundary of the half-space. The remaining fields $A_{z}$ and $X$ obey a generalized Neumann boundary condition\footnote{Note that, unlike \cite{Gaiotto:2008sa}, we use standard conventions for the $SU(2)$ algebra, $[t^i, t^j] = i \epsilon^{ijk} t^k$; our scalar fields correspondingly differ by a factor of $i$ from those of \cite{Gaiotto:2008sa}.}
\begin{equation}
\frac{DX^{i}}{Dz} + i \epsilon^{ijk}\, [X_{j}, X_{k}] = 0.
\label{nahm}
\end{equation}	
These boundary conditions arise for D3-branes ending on D5-branes (hence D-type).

\item Neumann or NS-type: This boundary condition involves Neumann boundary conditions for $A_{\mu}$ and for $Y$, whilst $X$ and $A_{z}$ satisfy Dirichlet boundary conditions. This boundary condition arises for D3-branes ending on a single NS5-brane. In this case the entire gauge symmetry of the $\CN=4$ SYM, $G$,  is preserved on the boundary of the half-space. As a result one has only $\CO(1)$ states in the spectrum at low energies in the large $N$ limit, owing to the singlet constraint discussed in \sec{s:n4ads}.
\end{itemize}

In the remainder of this section we will analyze the theory with these boundary conditions
in detail; more general boundary conditions will be considered in the next section.

\subsection{Nahm data for the D-type boundary conditions}
\label{s:nahm}

We will first try to understand some aspects of D3-branes ending on D5-branes, which will clarify the role of the generalized Neumann boundary conditions \req{nahm} relevant for the D-type boundary condition mentioned above. Recall that the D5-brane has a world-volume $U(1)$ gauge field that is sourced when a D3-brane ends on the five-brane; the end-point of the D3-brane is like a monopole in the D5-brane world-volume. A single D3-brane ending on a single D5-brane simply gives a Dirac monopole singularity at the intersection for the $U(1)$ gauge field. In order for multiple D3-branes to terminate on a single D5-brane one has to generalize the Dirac monopole solution; this is precisely achieved by  Nahm's equations \req{nahm}.

There is an intuitive way to think about this boundary condition and the solutions of Nahm's equations in terms of D-branes. In fact the solutions to Nahm's equation are the fuzzy funnel construction of \cite{Callan:1997kz}. As described there, the D3-branes ending on the D5-brane exert a force and distort the world-volume of the D5-brane. Alternately one can think of the D3-branes themselves flaring out to create a D5-brane as described in the non-commutative construction of \cite{Constable:1999ac}.

To see this more explicitly consider $N$ D3-branes that flare out as a fuzzy funnel and make up a single D5-brane. To obtain this by solving \req{nahm} we require that the $X$ fields provide a homomorphism from $SU(2)$ into $SU(N)$ which has a simple pole at $z=0$, i.e.,
\begin{equation}
X^i \xrightarrow{z\to 0} \frac{1}{z}\, t^i \ , \qquad \{t^i\} = \text{{\it N}-dimensional\ irreducible\ representation\ of\ } SU(2).
\label{xbc}
\end{equation}	
Since the scalar fields $X^i$ transform in this case under the $N$-dimensional representation of $SU(2)$, the $SU(N)$ gauge group of $ {\cal N} =4$ SYM is completely broken by this choice of boundary conditions. More generally, solutions to Nahm's equation can be given by prescribing a homomorphism from $SU(2)$ into ${\mathfrak g}$:
\begin{equation}
\rho : SU(2) \to {\mathfrak g} \ , \qquad X^i \xrightarrow{z\to 0} \frac{1}{z} \, t^i  ,\qquad  t^i \in \rho.
\label{xbcA}
\end{equation}	
Since the irreducible representations  of $SU(2)$ are classified by a positive integer, we just have to prescribe $k$ integers $n_i$ such that $\sum_{i=1}^k\, n_i = N$. Then the number of irreducible representations, $k$, of $SU(2)$ that occur tells us how many D5-branes are present. The conformal transformation
described above maps (\ref{xbcA}) to expectation values of the form $X^i = t^i / \ell_4$, which label
the supersymmetric vacua on \AdS{4} as described in \req{adssy} above.

A-priori one  might have expected that Dirichlet boundary conditions for the gauge field would imply that the entire gauge symmetry survives as a global symmetry on the boundary, with states carrying non-zero global charge allowed. As we have just seen, this however depends on  the choice of solutions to \req{nahm}, which allow singular configurations \req{xbcA} compatible with (and in fact required by) supersymmetry. These facts are not new and are well known in the context of D-brane intersections. In general for D$p$-branes ending on D$(p+2)$-branes one always has to make a choice of Nahm data as above \cite{Callan:1997kz}.

In the simplest case of a single D5-brane we are instructed to pick the $N$-dimensional irreducible representation of $SU(2)$. This choice of $\rho$ breaks the gauge group completely despite the Dirichlet boundary condition on $A_\mu$ and $Y$. Working directly in \AdS{4}, we give constant vacuum expectation values for the scalar fields to satisfy \req{adssy}, which break the gauge group in the bulk and give a mass to the vector fields; from the point of view of a putative dual CFT, the global symmetry group is explicitly broken by the boundary conditions.

In general the choice of map $\rho$ from $SU(2)$ to the gauge group $G$ could leave a subgroup $K$ as a global symmetry. This is easy to see in the brane construction where by picking $k$ irreducible representations of $SU(2)$, one constructs $k$ D5-branes out of the $N$ D3-branes. The world-volume gauge fields on the D5-branes act as global symmetries on the D3-brane fields; when the $k$ representations are identical we obtain an $SU(k)$ global symmetry.\footnote{Global symmetries arise from D5-branes only when there is more than one present; for $k=1$, there  is complete breaking of the global symmetries \cite{Gaiotto:2008sa}.}

In such cases we then are allowed to consider states carrying charge under $K$.  Depending on how large $K\subset G$ is, we could get a large number of  light charged degrees of freedom (which would contribute to the free energy).  As a special case consider $\rho$ being given by $N$ copies of the trivial representation of $SU(2)$, $t^i=0$. Then we construct $N$ D5-branes via the fuzzy funnel construction, and these give rise to a $SU(N)$ global symmetry.

With this understanding of the Nahm pole data relevant for the D-type boundary conditions, we are in a position to deconstruct the abstract data necessary to completely specify the boundary conditions for ${\cal N} =4$ SYM on the half-space. The data triple
$\left(\rho,H,{\mathfrak B}\right)$ can be interpreted as follows. Start with a gauge group $G$ (Lie algebra ${\mathfrak g}$) which defines the four dimensional ${\cal N} =4$ theory. Pick a map $\rho: SU(2) \to {\mathfrak g}$ which describes the Nahm data, essentially telling us the allowed pole structure of the scalars $X^i$ on the half-space, or equivalently the choice of a supersymmetric vacuum on \AdS{4}. The residual gauge symmetry after the choice of such a homomorphism is $H$, and we impose Neumann boundary conditions for $H$-valued fields, and we can furthermore couple them to a `boundary theory' ${\mathfrak B}$. So far we have not encountered any specific examples of ${\mathfrak B}$ -- this will be shortly remedied when we examine S-duals of the Dirichlet boundary conditions.

\subsection{S-duals of basic boundary conditions}
\label{s:sduals}

The role of the scalar VEVs (Nahm data) in the boundary conditions helps clarify an important point that caused some puzzle in the large $N$ limit, as discussed in the Introduction. In the case of Neumann boundary conditions we have the full $G$ gauge symmetry preserved on the boundary. As a consequence one expects a free energy of  $\CO(1)$ arising from the allowed gauge singlet states. Naively, based on the behaviour of the Abelian theory, one might have thought that the S-dual of the Neumann boundary condition is the Dirichlet boundary condition, with $G$ as the global symmetry in the latter. Were this to be true one would have predicted that there are $\CO(N^2)$ light states in the theory at large $N$ and strong coupling.

Of course, this is not really possible, since the S-duality cannot change the global symmetry group.\footnote{Note that in the Abelian theory there is a new $U(1)$ global symmetry arising in the
S-dual theory, with a current that is the Hodge-dual of the field strength on the boundary, but this does not happen in the non-Abelian case.} The scalar VEVs (Nahm poles) help to resolve this confusion;
in fact the S-dual of the Neumann boundary condition is the Dirichlet boundary condition with scalar VEVs breaking $G$ down to a trivial group \cite{Gaiotto:2008ak}. This is easy to see in the brane construction: the Neumann boundary conditions can be realized by ending the D3-branes on a single NS5-brane, and a S-duality transformation maps the D3-branes to themselves and the NS5-brane to a D5-brane. Since there is only a single D5-brane, one necessarily has scalar VEVs given by the $N$-dimensional irreducible representation of $SU(2)$, breaking the global symmetry completely. Despite the fact that the theory has no global symmetry, we do still have
many ``charged'' fields in the bulk (the W-bosons), which lead to a large number of degrees of freedom. However,
 all of these states are very heavy (with masses scaling as a positive power of $N$ in the large $N$
 limit) due to the Higgs mechanism resulting from \req{xbc}, so we do not get a large number of light states in the large $N$ limit, and S-duality does not change the qualitative low-energy behaviour of the theory. Note that
 for these boundary conditions, the fact that we have a small number of light degrees of freedom is due
 to ``confinement'' in the Neumann picture (albeit a kinetic confinement due to Gauss' law), and to
 Higgsing in the S-dual picture, in agreement with the general intuition that S-duality should
 exchange confinement in ``electric variables'' with the Higgs mechanism in ``magnetic variables''.

On the other hand, the Dirichlet boundary conditions preserving the entire global symmetry are obtained by having $N$ D3-branes ending on $N$ D5-branes  (or more generally picking trivial homomorphisms $\rho$ for general ${\mathfrak g}$). The S-dual is then given by $N$ D3-branes ending on $N$ NS5-branes. This is a non-trivial boundary condition, and for the first time we encounter the third ingredient mentioned in the general supersymmetric boundary conditions earlier, viz., the boundary CFT ${\mathfrak B}$. This theory is a 2+1 dimensional CFT with ${\cal N} = 4$ supersymmetry. It was shown in \cite{Gaiotto:2008ak} that this theory ${\mathfrak B}$ is a quiver type theory called $T(G)$, and it has $G \times G^\vee$ global symmetry, with $G^\vee$ being the dual of $G$.

The theory $T(SU(N))$ may be described in terms of branes. We consider $N$ NS5-branes ordered along the direction $z$, with $j$ D3-branes between the $j^{\rm th}$ and $(j+1)^{\rm st}$ NS5-brane, and with the leftmost NS5-brane having a single D3-brane ending on it. The $T(SU(N))$ theory is the superconformal $2+1$ dimensional CFT which arises in the infra-red limit of such a quiver gauge theory; in this limit the NS5-branes overlap, so
we get an $SU(N)\times SU(N)$ global symmetry from the $N$ NS5-branes and the $N$ D3-branes going out of the intersection region. In our construction above the D3-brane factor of the global symmetry is gauged, while the NS5-brane factor is S-dual to the original $SU(N)$ global symmetry associated with the Dirichlet boundary conditions. In any case, on both sides of S-duality we find a large number of light degrees of freedom
in this case.

\begin{table}[t]
\begin{center}
\begin{tabular}{|c|c||c|}
\hline
{\bf Boundary condition} &{\bf Realization}& {\bf Dual boundary condition}
\\\hline \hline
&&\\
Neumann for $SU(N)$ & $N$ D3-branes ending & Dirichlet with Nahm pole\\
& on a NS5-brane& $ \rho = {\bf N}$-dimensional irrep. \\
&&
  \\\hline
  && \\
Dirichlet for $SU(N)$ & $N$ D3-branes ending  & $(3+1)$-dim ${\cal N} =4$ SYM coupled to\\
&on $N$ D5-branes&  $(2+1)$-dim $T(SU(N))$ quiver  \\
&&
  \\\hline
 \end{tabular}
 \caption{The basic boundary conditions and their duals. The field theories are engineered by  taking $N$ D3-branes lying along $(0123)$ with $z$ being restricted to $z\ge 0$. D5-branes are taken to lie along $(012456)$ while NS5-branes occupy $(012789)$. $SU(2)$ representations are labeled by appropriate integers, with ${\bf 2j \!+\!1}$ denoting the spin $j$ representation.}
\label{t:basicd}
\end{center}
\end{table}

More generally, the strategy for finding the S-dual of a given boundary condition $ \left(\rho, H,{\mathfrak B}\right)$ is described in detail in \cite{Gaiotto:2008ak}. For the case of unitary gauge groups having brane constructions, one can use the fact that NS5-branes and D5-branes are exchanged by S-duality. This has to be further supplemented by some brane translations to be able to read off the dual boundary condition. We will not review here the precise strategies for recovering this dual data, since we will see momentarily that even the simplest cases are not amenable to holographic treatment at strong coupling.

\subsection{Holographic descriptions of basic boundary conditions}
\label{s:holimp}

Having resolved our puzzle regarding the nature of excitations of ${\cal N} =4$ SYM on \AdS{4} with Neumann and Dirichlet boundary conditions at very strong coupling $g_{YM} \gg 1$, we would next like to understand the holographic duals  of the theory with these boundary conditions. To keep things simple we will only describe the story for unitary gauge groups, which will allow us to use the language of branes in string theory.  Generically whenever we have a global symmetry group $\tilde{G}$ in our field theory, this should be realized as a gauge symmetry in the bulk of the gravitational dual. In principle, for a theory described as the low-energy limit of D3-branes ending on some brane configuration, all we need to do to find the holographic dual is to find the gravitational solution describing the back-reaction of the D3-branes, and take their near-horizon limit. Note that the global symmetries in brane constructions arise from gauge symmetries on the branes that the D3-branes end on; these branes are expected to be present in the near-horizon limit and to provide the required gauge fields in the bulk. Unfortunately, no gravitational solution for D3-branes ending on other branes is known (though solutions for localized intersecting branes were found in \cite{DHoker:2007xy,DHoker:2007xz}), so we do not have any known solutions to work with, and we can only discuss the qualitative features that we expect the solutions to have.

Let us start with Dirichlet boundary conditions for ${\cal N} =4$ SYM on \AdS{4}, described in
\sec{s:gaugeads4}. As mentioned above, the fact that these boundary conditions for the theory on \AdS{4} have ${\cal O}(N^2)$ light bulk excitations (e.g., from the gauge fields dual to the global symmetry currents) poses a strong challenge for the holographic duals. We saw above that
on the half-space this boundary condition is realized by having extra branes in the game. In particular, for $G = SU(N)$ and Dirichlet boundary conditions that do not break $G$, we require $N$ D5-branes in addition to the $N$ D3-branes that conjure up the ${\cal N} =4$ bulk dynamics in the theory.  If the number of D5-branes was small compared to the D3-branes, say $k$ with $k \ll N$, and if only a small fraction of the D3-branes would end on them, then one would be able to treat the D5-branes in a probe approximation to leading order. This analysis would reveal that the world-volume of the D5-branes is an \AdS{4} $\times\, {\bf S}^2$ $\subset$ \AdS{5} $\times\, {\bf S}^5$ (see, e.g., \cite{Karch:2000gx}). However, in our case we have a large number of D5-branes with a strong back-reaction, and moreover we
expect these D5-branes to provide an end for the \AdS{4} space-time on the boundary, so the
probe analysis is clearly irrelevant. As mentioned above, it seems that the dynamics of the gravitational dual in this case (at least of the bulk $SU(N)$ gauge symmetry) is strongly coupled, so it is unlikely that a weakly coupled and curved gravitational dual exists (at least not in the region near the boundary, where we know we should have the $SU(N)$ gauge fields). One may be tempted to replace the $N$ D5-branes by some gravitational dual, but this throws away all $SU(N)$ non-singlets, and does not seem satisfactory since we should be able to describe arbitrary sources for the $SU(N)$ currents (and other non-singlets). Note that whatever the dual background is, its S-dual is related to $N$ D3-branes ending on $N$ NS5-branes, which as we have discussed above corresponds to the ${\cal N} =4$ theory on \AdS{4} coupled to the $2+1$ dimensional $T(SU(N))$ theory living on its boundary \ESU{3}.

Next we discuss Dirichlet boundary conditions with scalar VEVs (Nahm poles), where the precise details depend now on the homomorphism $\rho: SU(2) \to {\mathfrak g}$. Again for simplicity we focus on unitary $G$, where such boundary conditions can be realized by $N$ D3-branes ending on a smaller number, say $k$, of D5-branes. This can be done in many ways, preserving some subgroup $H$ of $G$ as the global symmetry.  At one extreme we could ensure that the full $G$ symmetry is broken by the scalar VEVs required to construct the D5-branes. An explicit example with $k=1$ is provided when $G = SU(N)$ and one takes $\rho$ to be the $N$-dimensional irreducible representation as in \req{xbc}.

In such cases we have just a few D5-branes and a small global symmetry (at most $SU(k)$ for $k$ D5-branes). Could these examples have holographic duals in terms of weakly curved backgrounds ? Take for example the case of $k=1$: we know that in this case the S-dual is the Neumann boundary condition attained by ending $N$ D3-branes on a single NS5-brane. In this description we do not have any scalar VEVs, just the source provided by the NS5-brane.

Unfortunately the solution for $N$ D3-branes ending on a single D5-brane or NS5-brane is not known. The fivebrane should certainly have a significant back-reaction -- in particular it should create a boundary for the boundary of \AdS{5}. Note that even though naively a single D5-brane does not have a large back-reaction at weak coupling, this is not true when the D5-brane carries $N$ units of D3-brane flux (as it does when $N$ D3-branes end on it). For the NS5-brane boundary at weak coupling we noted that the number of degrees of freedom at low energies was of order one, so it is plausible that this case could have a smooth holographic dual. The case of a D5-brane boundary is more confusing, since in this case at weak coupling the Higgs mechanism implies that there are no light fields in the large $N$ limit,\footnote{We know this to be true at small 't Hooft coupling, but it is unlikely to change at finite 't Hooft coupling, since all masses scale as positive powers of $N$.}
 and presumably no sensible holographic dual as well. Since this case is related by S-duality to the single NS5-brane case, the holographic dual of the  latter must be somewhat peculiar (so that under
S-duality it becomes singular, or at least harbors no light states). It seems more likely that the
theory related to $N$ D3-branes ending on $k$ NS5-branes or $k$ D5-branes, with $k > 1$ fixed in the large $N$ limit, could have a smooth holographic dual, since from the NS5-brane point of view this
allows the NS5-branes to be replaced by a smooth ``throat'', and from the D5-brane point of view
this leaves some light degrees of freedom that are not Higgsed at weak coupling.

We can also think about other boundary conditions. The general boundary condition of interest can be described as a Neumann boundary condition for a subgroup $H \subset G$. As a result, on the boundary of the half-space and hence on the boundary of \AdS{4}, we have a surviving gauge symmetry for $H$-gauge fields. This means that, for large enough $H$, we should expect $\CO(1)$ states in the theory at low temperatures due to Gauss' law, and there is a chance of having a smooth holographic dual. If $H$ happens to be small, we would get little  from Gauss' law, and as a result we would still have $\CO(N^2)$ states in the spectrum. These cases are similar to our discussion above of the Dirichlet type boundary condition, and again we may expect a large number of branes in the bulk complicating any holographic analysis.

\section{$\CN=4$ SYM on \AdS{4} with quotient boundary conditions}
\label{s:bcquotients}

The boundary conditions we discussed so far are the simplest ones; more complicated boundary conditions were also described in \cite{Gaiotto:2008sa} and we will next review these. To understand the rationale for this, note that one of our motivations is to understand the boundary conditions which allow us to study dynamics on \AdS{4} at strong coupling using holographic methods. We have already seen that the simplest versions of the Dirichlet  boundary conditions are not amenable to a useful gravitational dual, and while it is possible that other boundary conditions like Neumann have a useful dual, new gravity (or string theory)
solutions need to be found to confirm this.

In order to generate boundary conditions that are amenable to a simple holographic treatment, we need to add additional ingredients to the story. We will now argue that this can be achieved by boundary conditions involving quotients;  in string theory this amounts to allowing orbifold/orientifold five-planes. A simple argument to motivate the necessity of the quotients we describe below is to realize that the canonical \AdS{4} foliation of \AdS{5} \req{adsinads}, described above, has two \AdS{4} boundary components. We want to describe the field theory on a single \AdS{4}, so we need a reflection that identifies these two `boundary \AdS{4}'s. To preserve supersymmetry the quotient has to also act on the transverse space ($\Sp^{5}$ for the case of $\CN=4$ SYM). Thus by a suitable $\ZZ_2$ quotient we can use known supergravity solutions to gain some insight into the strong coupling limit of
${\cal N} =4$ SYM on \AdS{4}, albeit at the price of introducing more exotic sounding boundary conditions.

 Another reason for considering more exotic boundary conditions which are a quotient construction follows from the original treatment of  \cite{Gaiotto:2008ak}. The simplest $D$ or $NS$-type boundary conditions, described in \sec{s:bc2}, demand that on the half-space the fields obey a prescribed boundary condition as $z \to 0$. However, one can also get the half-space ${\bf R}^{2,1} \times {\bf R}_+$ from ${\bf R}^{3,1}$ by a  $\ZZ_2$ involution which reflects $z \to -z$. This means that one can use the usual folding trick to obtain new boundary conditions. In general the $\ZZ_2$ quotient also acts on the gauge group $G$, and the choice of boundary conditions depends on the precise nature of this action.

The quotients we will describe below will turn out to leave an \AdS{4} $\times \ \Sp^{2}$ fixed hypersurface inside \AdS{5} $\times\ \Sp^{5}$. For simplicity we will again describe these quotients in the half-space version as described in \cite{Gaiotto:2008ak}. As before, for ease of illustration, we will use the familiar language of D-branes, orientifold planes and world-sheet parity actions.

\subsection{Boundary conditions with involutions}
\label{s:bcinvo}

Let us again start with ${\cal N} =4$ SYM on the half-space with gauge group $G$. We want to preserve a sub-group $H \subset G$ and impose Neumann boundary conditions on gauge fields in $H$. As explained in \cite{Gaiotto:2008sa}, we can start with the theory on ${\bf R}^{3,1}$ and define the theory on the half-space by the folding trick, obtaining the half-space by the quotient $z \to -z$. In order to preserve supersymmetry, the $\ZZ_2$ action should also act by reflecting three of the scalars of ${\cal N } =4$ SYM (this amounts to reflection of four directions in string theory).

Now let us pick some $\ZZ_2$ grading of the Lie algebra $\mathfrak g$. We can write the corresponding Lie algebra decomposition as ${\mathfrak g} = {\mathfrak h} +{\mathfrak h}^\perp$, where $H$ is a subgroup of $G$. All fields of the theory (collectively denoted $\Psi$) can be split into those that take values in ${\mathfrak h}$, $\Psi_{\mathfrak h}$, and those that live in the complement ${\mathfrak h}^\perp$, $\Psi_{{\mathfrak h}^\perp}$. Under the $\ZZ_2$ action we can then assign charge $+1$ to the fields $\Psi_{\mathfrak h}$ and charge $-1$ to $\Psi_{{\mathfrak h}^\perp}$; this just amounts to supplementing the reflections with an involution of the gauge group.

The analysis of the supersymmetry transformations reveals that it is possible to give Neumann boundary conditions to the $\Psi_{\mathfrak h}$ fields. Thus the part of the gauge symmetry that is preserved on the boundary is just $H$. For the most part we will now focus on the case of unitary gauge groups $G = SU(N)$.  Since these can be realized as the world-volume theory on D3-branes, we can phrase the discussion in terms of the string theoretic constructions.

To understand what reductions can be achieved, one has to classify involutions of $SU(N)$; a description of these can be found in \cite{Gaiotto:2008ak}, and they are summarized below:
\begin{itemize}
\item Class I involution: conjugation by an element that breaks $SU(N) \to SO(N)$.
\item Class II involution: conjugation by an element that breaks $SU(N) \to USp(N)$.
\item Class III involution:  conjugation by an element that breaks $SU(N) \to SU(p) \times SU(N~-~p)\times U(1)$ (we can always choose $p \ge \frac{N}{2}$, and we will mostly ignore the $U(1)$ factor).
\end{itemize}
The involutions of Class I and II are outer automorphisms of the gauge group, while the Class III involution is an inner automorphism.  By utilizing the above choices of involutions one can engineer boundary conditions where the gauge symmetry $G$  is broken down to $H$ which can be one of $\{ SO(N), USp(N), SU(p) \times SU(N-p)\}$ depending on the choice made. There may be further breaking of the gauge symmetry if we choose a vacuum with non-trivial scalar VEVs, and moreover we can further consider coupling the four dimensional theory to some $2+1$-dimensional CFT with $H$ symmetry. However, the simplest choice of boundary conditions involves no scalar VEVs or boundary CFT; one simply uses the involution to gauge (on the boundary) a subgroup of the gauge symmetry $G$.

The aforementioned quotients have a nice interpretation in string theory. As before we will engineer boundary conditions using D3, D5 and NS5-branes. In addition since we are performing quotients, we can include orbifolds and orientifolds. It turns out that given the supersymmetries we wish to preserve, one has two choices:
\begin{itemize}
\item An orbifold by ${\cal I}_4 \, (-1)^{F_{L}}$. Here ${\cal I}_{4}$ is a spacetime reflection on four directions, three directions transverse to the D3-brane along with $z \to -z$, and $(-1)^{F_L}$ is the operator that counts world-sheet left-moving fermions.
\item An orientifold which involves the same ${\cal I}_4$ reflection as above along with a world-sheet orientation reversal.
\end{itemize}

The quotient constructions in string theory have a natural mapping to the involutions of unitary $G$ mentioned above. The orbifold by ${\cal I}_4 \, (-1)^{F_{L}}$ has to be supplemented with an action on the Chan-Paton factors. One has a choice of a discrete $\ZZ_2$ action by way of assigning charges $\pm 1$ to the Chan-Paton factors. Since one can pick, say, $p$ D3-branes where the action is $+$ and $q$ with action $-$, the orbifold realizes the Class III involution, thereby breaking the gauge group down to $ SU(p) \times SU(q)\times U(1)$ with $p+q = N$. It is also useful to note that this orbifold has a twisted sector $U(1)$ gauge field that lives on the fixed plane.

In the case of the orientifold, the quotient itself breaks $SU(N)$ down to $USp(N)$ or to $SO(N)$ depending on the choice of orientifold plane ($O5^{\mp}$, respectively). Apart from the five-brane charge carried by the orientifolds, the two choices of $O5^\pm$ are distinguished by the flavour symmetry acting on the D3-brane fields: $m$ D5-branes coincident with an $O5^-$ have $SO(2m)$ gauge symmetry, while those atop an $O5^+$ have a $USp(2m)$ gauge symmetry. For a set of D3-branes which end on such D5+$O5$, these gauge symmetries of the D5-brane theory act as flavour symmetries on the $2+1$-dimensional hyper-multiplet fields living on the boundary, coming from D3-D5 strings.
The flavour symmetry is useful to relate the two choices of the orientifold action with the Class I and II involutions described above.  The Class I involution which breaks $SU(N) \to SO(N)$ involves an $O5^+$ orientifold, while the Class II involution breaking $SU(N) \to USp(N)$ is related to an $O5^-$ orientifold plane. A summary of the basic constructions is provided in Table \ref{t:quosum} for quick reference.

\begin{table}[t]
\begin{center}
\begin{tabular}{|c|c|c|}
\hline
{\bf Involution} &{\bf Quotient}& {\bf Gauge group $H\subset G$ }
\\\hline \hline
Class I &  $O5^{+}$ & $SO(N)$ \\ \hline
Class II & $O5^-$ & $USp(N)$ \\ \hline
Class III &  ${\cal I}_{4} (-1)^{F_{L}}$ & $SU(p) \times SU(N-p)$  \\\hline
 \end{tabular}
 \caption{A summary of the involutions breaking the gauge symmetry and their realization in string theory. }
\label{t:quosum}
\end{center}
\end{table}

As before, these simple ingredients do not give all possible boundary conditions; one can also have scalar VEVs which reduce the symmetry, and/or explicit breaking of symmetry by further Neumann boundary conditions. Rather than explain these in abstraction, we will introduce them as necessary when we describe the action of the S-duality between these various quotient constructions momentarily.

\subsection{S-duals of the quotients}
\label{s:involsd}

In order to understand the role of S-duality, we first note that the orientifolding action  can be chosen to preserve the same supersymmetries as a D5-brane. In fact, the orientifolds carry D5-brane charge, putting them on the same footing as the D5-branes. The orbifold by ${\cal I}_4 \, (-1)^{F_L}$ on the other hand has no net five-brane charge; nevertheless it is similar to (and preserves the same supersymmetries as) an NS5-brane in type IIB string theory.
Thus, one expects that the action of the S-duality will be to exchange the orbifold and orientifold boundary conditions. A basic summary of the results from \cite{Gaiotto:2008ak} is outlined in Table \ref{t:orborient}.

One way to proceed is to note the five-brane charges involved. Realizing that the orbifold by ${\cal I}_4 \, (-1)^{F_L}$ has zero five-brane charge tells us that its S-dual can be given in terms of the $O5^-$ orientifold, provided that we supply a D5-brane on the orientifold locus to ensure vanishing charge. This identification is also supported by the flavour symmetries; we have already noted that the twisted sector of the orbifold supports a $U(1)$ gauge field (which acts as a flavour symmetry on the D3-brane hypermultiplets). A single D5-brane coincident with the $O5^-$ plane also has a $SO(2) \simeq U(1)$ gauge symmetry on the D5-brane world-volume, which acts as a flavour symmetry on the boundary hypermultiplets.

However, this naive S-duality identification between the $O5^-$ orientifold (with a D5-brane) and the ${\cal I}_4 \, (-1)^{F_L}$ orbifold leaves something to be explained. In the case of the $O5^-$ orientifold, there is a unique gauge group obtained by the action of the involution, $USp(N)$. On the other hand, the orbifold action corresponds to the Class III involution, and therefore to a discrete set of theories labeled by a single integer $p$. This implies that more ingredients are required to describe the S-dual of the orbifold. In fact, the extra ingredients are nothing but scalar VEVs (Nahm poles). In particular, as argued in \cite{Gaiotto:2008ak}:
\begin{itemize}
\item The simplest case is $p = N/2$, where the orbifold action is symmetric and the resulting gauge group is $SU(N/2) \times SU(N/2)$.  The  S-dual of these boundary conditions is an orientifold $O5^-$ plane together with a single D5-brane stuck at the fixed plane. This S-dual theory has extra boundary degrees of freedom given by hypermultiplet fields which are charged under $USp(N)$ and under the $SO(2) \simeq U(1)$ flavour symmetry, arising from the D3-D5 strings.
\item For general $p$, the orbifolded theory has gauge symmetry $SU(p) \times SU(q)$ with $p+q = N$.  The S-dual for $p \ge  q+2$ involves a Nahm pole $\rho$ which breaks $SU(N)$ down to $SU(2q)$, combined with Neumann boundary conditions reducing the gauge group down to $USp(2q)$. This is achieved by a map $\rho: SU(2) \to SU(N)$ which is composed of the dimension $(p-q)$ irreducible representation together with $2q$ copies of the trivial representation. This Nahm pole preserves a $SU(2q) \subset SU(N)$ gauge symmetry; a subsequent symmetry breaking boundary condition,  imposing a Neumann boundary condition for $USp(2 q) \subset SU(2q)$, results in a $USp(2q)$ gauge theory. The case $p=q+1$ simply involves symmetry breaking boundary conditions without a Nahm pole, the symmetry group being just $USp(N-1)$. In this generic case of $p\neq q$ one does not require D5-brane degrees of freedom (these are effectively provided by the scalar VEVs).
\end{itemize}
Thus the S-duals of the orbifold boundary conditions all involve modifications of the $O5^-$ orientifold (either by explicit introduction of the D5-brane for $p=q$, or by Nahm poles for general $p\neq q$).

\begin{table}[t]
\begin{center}
\begin{tabular}{|c|c||c|c|c|}
\hline
{\bf Gauge group} &{\bf Quotient}& {\bf Dual gauge } & {\bf Nahm data} & {\bf Boundary}\\
 $H$& & {\bf group ${\widetilde H}$} & {\bf for dual}& {\bf data (dual)}
\\\hline \hline
&&&&  D5-brane \\
$SU(\frac{N}{2})\times SU(\frac{N}{2})$ & ${\cal I}_{4} (-1)^{F_{L}}$ & $USp(N)$ & $\rho = N\times {\bf 1}$  & localized \\
&&&& on $O$-plane
  \\\hline
 &&&&  Symmetry\\
  $SU(p) \times SU(q)$ &${\cal I}_{4} (-1)^{F_{L}}$&  $USp(2q)$ & $\rho = ({\bf p\!-\!q} )\oplus 2q \times {\bf 1} $& breaking \\
   &&&& Nahm data
  \\\hline \hline
  & & & & Non-trivial\\
$SO(N)$ & $O5^{+}$ & $SU(N)$ & $\rho =N\times {\bf 1 }$& $2+1$d SCFT\\
&&&& on boundary
\\ \hline
&&&& \\
 $USp(N)$ & $O5^{-}$ & $SU\left(\frac{N}{2}\right)_{d} \subset SU(N)$ &   $\rho = \frac{N}{2}\times {\bf 2}$& None\\
 &&&&
  \\\hline
 \end{tabular}
 \caption{The possible gauge theories engineered by the orbifold/orientifold action and their S-dual partners. The first two columns refer to the gauge theory obtained by the quotient action, and the last three columns provide the data describing the S-dual theory. The notation is as in Table \ref{t:basicd}, with orientifold planes along $(012456)$ and the orbifold plane along $(012789)$. Here $SU(\frac{N}{2})_d$ denotes the diagonal subgroup in the decomposition $SU(\frac{N}{2}) \times SU(\frac{N}{2}) \subset SU(N)$. Note that the S-dual of a quotient in general involves action of an involution and in addition some Nahm data and/or extra boundary degrees of freedom.  }
\label{t:orborient}
\end{center}
\end{table}

Having understood the S-duals of the orbifold constructions, we now turn to the orientifold boundary conditions. There are  two cases to consider depending on the five-brane charge of the $O5$-plane:
\begin{itemize}
\item The $O5^+$ orientifold (Class I involution) breaks the $SU(N)$ of the D3-branes to $SO(N)$. This orientifold is S-dual to some configuration of D3-branes ending on a  NS5-brane (which follows from the fact that the D5-brane charge of the $O5^+$ dualizes to the NS5-brane charge). However, the boundary condition in the dual cannot simply be Neumann boundary conditions for the $SU(N)$ (whose S-dual we already discussed in \sec{s:sduals}), and the S-dual involves coupling  to a non-trivial field theory living on the boundary (Nahm poles are forbidden since the D5-brane charge vanishes for the dual).
\item The $O5^-$ orientifold is the Class II involution which leaves a residual $USp(N)$ gauge symmetry on the D3-branes. We have encountered variants of this boundary condition in our discussion of the duals of the orbifold boundary conditions. However, in the present case we require the dual of just the orientifold plane. As described in \cite{Gaiotto:2008ak}, the dual boundary condition is simply given in terms of scalar VEVs.  These break $SU(N) \to SU(N/2)_d \subset SU(N/2) \times SU(N/2)$ in the S-dual picture, and involve $N/2$ copies of the doublet representation of $SU(2)$, and Neumann boundary conditions for $SU(N/2)_d$. There are no additional boundary degrees of freedom in the description.
\end{itemize}

As mentioned above, the S-duals of the simple orbifold and orientifold boundary conditions involve extra ingredients: scalar VEVs, boundary symmetry breaking and boundary SCFT. For the basic quotients we have provided a summary of the results from \cite{Gaiotto:2008ak}, in a form that will be suitable to our investigations. We should note  that the discussion can be enlarged to include quotients of orthogonal and symplectic gauge groups (which in string theory involve $O3$-planes as well), and to also allow extra boundary degrees of freedom. We refer the interested reader to \cite{Gaiotto:2008ak} for details on these constructions.

\subsection{${\cal N} =4$ SYM with quotient boundary conditions at weak coupling}
\label{s:involweak}

Having understood the possible boundary conditions for ${\cal N} =4$ SYM involving symmetry breaking involutions, we can now investigate the theory on \AdS{4} at weak coupling. (Recall that  it is simple to conformally map the boundary conditions from the half-space into \AdS{4}, where the Nahm data is interpreted as the choice of a supersymmetric vacuum.) We have three basic classes of theories to deal with, depending on the choice of the involution.  In each case, as described in \sec{s:bcinvo}, we break the gauge symmetry $G =SU(N)$ down to a subgroup $H$. We would like to ascertain the spectrum of the theory at weak coupling. Since $H$ has a rank of order $N$ in all of the cases, we expect an interesting phase structure for Neumann boundary conditions, involving a Hagedorn transition at a temperature set by the \AdS{4} length scale.

The basic result we need has already been obtained in \req{HNeupf} for the symmetry breaking from a gauge group $G$ to $H$. We will now explore this result in the concrete setting of the quotients explained above, where $G = SU(N)$ and we have symmetry breaking patterns to $H = \{ SU(p) \times SU(q), SO(N), USp(N)\}$.

Essentially all we need to make explicit the formula \req{HNeupf} are the branching rules for the adjoint representation of $G= SU(N)$ into appropriate representations of
$H$,  i.e.,
\begin{equation}
\text{adj}(SU(N)) = {\bf N^2\!-\!1} = \sum_i \,m_i \, R_i\,,
\label{}
\end{equation}	
where $m_i$ are the multiplicities with which the representation $R_i$ of $H$ occurs. For the quotients of interest the decomposition of the representations is:
\begin{eqnarray}
SU(p) \times SU(q) \hookrightarrow SU(N) : && {\bf N^2\!-\!1} =
\left({\bf p^2\!-\!1  },{\bf 1 }\right) \oplus \left({\bf 1},{\bf q^2\!-\!1}\right) \oplus \left({\bf p },{\bf \bar{q}}\right) \oplus \left({\bf \bar{p}},{\bf  q}\right) \oplus \left({\bf 1}, {\bf 1}\right)  \nonumber \\
&& \nonumber \\
SO(N) \hookrightarrow SU(N) : && {\bf N^2\!-\!1} =
{\bf \frac{N(N-1)}{2}}\oplus \left({\bf \frac{N(N+1)}{2} -1} \right) \nonumber \\
&& \nonumber \\
USp(N) \hookrightarrow SU(N)  : &&  {\bf N^2\!-\!1}  =
\left({\bf \frac{N(N+1)}{2} -1} \right) \oplus {\bf \frac{N(N-1)}{2}}
\label{branching}
\end{eqnarray}	
For the unitary subgroup $H = SU(p) \times SU(q)$ we get adjoint representations for each component along with bi-fundamentals and a singlet. The orthogonal subgroup $H=SO(N)$ comes with the adjoint (rank $2$ anti-symmetric tensor) and the symmetric, traceless 2-tensor representation. The symplectic subgroup $USp(N)$ likewise has matter in the rank 2 anti-symmetric tensor representation in addition to the adjoint gauge multiplet.

The computation of the large $N$ Hagedorn temperature from \req{HNeupf} can be done using the techniques described in \cite{Aharony:2003sx}, which we review in \App{s:NeuHag}. The upshot of the calculation is that for the three classes of quotients described above (assuming for the Class III involution that both $p$ and $q$ are of order $N$)\footnote{For the Class III involution when $p \sim N$ but $q \ll N$,  one has to be a bit more careful since the eigenvalue distribution for the $SU(q)$ gauge group cannot be treated in a continuum approximation.  In fact, for $p=N-k$ with finite $k$ one expects that the large $N$ transition temperature is given by \req{weakHagA}.} one finds a transition temperature:
\begin{equation}
T_\star = -\frac{1}{\ell_4\, \log (x_\star)} \ , \text{with} \,\, z(x_\star) = z_B(x_\star) + z_F(x_\star) = \frac{1}{2}.
\label{Hagquo}
\end{equation}	
The factor of $1/2$ on the right-hand side of (\ref{Hagquo}) can be simply understood from the fact that, as implied by the
previous paragraph, when we construct a single-trace operator we always have twice as many choices for
which field to put at each position compared to the original unquotiented theory.
For all the quotients of interest, the Hagedorn temperature (taking $z_B(x)$ with $\alpha =3$ as required by supersymmetry) is found to be $T_\star = 1 / (\ell_4\, \log(7 + 4\,\sqrt{3}))$.

\subsection{Holographic duals for the quotient constructions}
\label{s:holoA1}

 Next, we wish to understand the holographic duals of $\CN=4$ SYM on \AdS{4} for the class of quotient boundary conditions described in \sec{s:bcinvo}.
Fortunately, we are in luck; we can find a holographic dual with a single copy of \AdS{4} on its boundary by taking a quotient of \req{adsinads} which identifies the two copies of \AdS{4} on the boundary of the \AdS{5} spacetime. That such an identification exists and is consistent in the full string theory follows from the discussion in \sec{s:bcinvo}. In particular, we require a $\ZZ_2$ involution which acts on \AdS{5} $\times {\bf S}^5$ as $R \to -R$ together with an action on the ${\bf S}^5$, combined perhaps with a further action on the string world-sheet.  This is precisely provided by the ${\cal I}_4\, (-1)^{F_L}$ orbifold and the orientifold constructions. As discussed above, these boundary conditions lead to $\CO(1)$ low-energy degrees of freedom, consistent with a smooth gravitational dual.

Let us focus on the simple case of the $O5^-$ orientifold, which breaks the $SU(N)$ gauge symmetry on $N$ D3-branes down to $USp(N)$. The dual geometry is \AdS{5}$ \times \ {\bf S}^5$ up to identifications. The orientifold action involves mapping the two copies of the \AdS{4} boundary in the coordinatization \req{adsinads} to each other, and it also acts by reversing the sign of three of the coordinates in the ${\bf R}^6 \supset {\bf S}^5$. Writing the metric on ${\bf S}^5$ as $ds^2 = d\theta_1^2 + \sin^2(\theta_1) \, d\Omega_2^2  + \cos^2(\theta_1)\, d{\widetilde \Omega}_2^2$, the action can be geometrically described as $(R, \theta_1) \to (-R, -\theta_1)$ together with the world-sheet parity reflection.  These identifications are realized by an  orientifold 5-plane localized on an \AdS{4} $\times \ {\bf S}^2$ surface $(R=\theta_1=0)$ in \AdS{5} $\times {\bf S}^5$. In the supergravity limit with $g_s \sim \frac{1}{N} \ll 1$, the orientifold plane has a negligible backreaction. Thus  we have constructed a suitable dual for these orientifold boundary conditions. Similar considerations extend to the $O5^+$ orientifold which preserves a $SO(N)$ gauge symmetry, and the orbifold boundary conditions preserving $SU(p) \times SU(N-p)$ gauge symmetry. In the
latter case, the value of $p$ is related to the flux of the twisted sector gauge field on the ${\bf S}^2$.

\subsection{Holography with quotient boundary conditions at finite temperature}
\label{s:holoA}

Let us now ask whether the class of geometries considered above, the \AdS{4} foliations of \AdS{5}, provide us with some insight into the dynamics of strongly coupled $\CN=4$ SYM on \AdS{4} at finite temperature. The aim will be to ascertain the phase structure of the theory from the holographic description and compare this with the free field analysis above.

It is easy to identify one class of geometries relevant to the dynamics of thermal $\CN=4$ SYM on \AdS{4}. This is simply \req{adsinads} (with the appropriate $\ZZ_2$ quotient), where we identify the Euclidean time coordinate $t_E = -i\, t$ with a period set by the temperature $t_E = t_E + \beta$. Note that the coordinates employed in \req{adsinads} identify the bulk and boundary time coordinates, so thermal boundary conditions can be easily imposed by the usual Euclidean period.

However, there is another class of spacetimes which can also be potential duals. To understand these geometries let us start with \AdS{d+1} (for generality) written in global coordinates as in \req{global2}
\begin{equation}
ds^2 = -f(\rho) \, dt^2  + \rho^2 \, d\Omega_{d-1}^2 + \frac{d\rho^2}{f(\rho)}\,,
\label{adsglobal}
\end{equation}	
where we now use $\rho$ to denote the bulk radial coordinate, and write the metric on the round $\Sp^{d-1}$ as
\begin{equation}
d\Omega_{d-1}^2 = d\zeta^2 + \sin^2 (\zeta)\, d\Omega_{d-2}^2\,.
\label{sphmet}
\end{equation}	
The function $f(\rho)$ for static, spherically symmetric spacetimes is of the form:
\begin{equation}
f(\rho) = \frac{\rho^2}{L_{d+1}^2} + 1 - \frac{\rho_+^{d-2}}{\rho^{d-2}}\left( 1 + \frac{\rho_+^2}{L_{d+1}^2}\right),
\label{fdef}
\end{equation}	
with $\rho_+ =0$ corresponding to pure \AdS{d+1}, while generic values of $\rho_+$ are the \SAdS{d+1} solutions. As expected this geometry has a boundary which is the \ESU{d} parameterized by $\{t,\zeta, \Omega_{d-2}\}$.

We are going to exploit the fact that \AdS{d} is conformal to \ESU{d}. This can be used to  motivate a diffeomorphism of the bulk spacetime \AdS{d+1} which acts as a boundary conformal transformation, thereby allowing us to construct bulk geometries which have \AdS{d} as their boundary. The coordinate transformation in question is rather simple; we just  invert the map used to obtain \req{globalesu} from \req{global2}.  The coordinate transformation
\begin{equation}
\tan (\zeta) = \frac{r}{\ell_{d}}\ , \qquad {\bar \rho} =\frac{L_{d+1}}{\ell_d} \, \rho
\label{ea1}
\end{equation}	
applied to  \req{adsglobal} results in the metric
\begin{equation}
ds^2 = \frac{{\bar \rho^2}}{L_{d+1}^2 \, (1+\frac{r^2}{\ell_d^2})} \left(-\frac{f({\bar\rho})}{{\bar\rho}^2} \, L^2_{d+1}\, \left(1+\frac{r^2}{\ell_d^2}\right) \, dt^2 + \frac{dr^2}{(1+\frac{r^2}{\ell_d^2})} + r^2\, d\Omega_{d-2}^2 \right) + \frac{\ell_d^2}{L_{d+1}^2} \, \frac{d{\bar \rho}^2}{f({\bar\rho})}.
\label{warpads}
\end{equation}	
We have now  achieved the stated goal of writing $d+1$-dimensional static, spherically symmetric asymptotically globally \AdS{d+1} spacetimes in coordinates where we have an \AdS{d} boundary (up to a conformal transformation, and the need for a $\ZZ_2$ quotient as discussed above).

The metric \req{warpads} with $d=4$ provides a class of potential dual spacetimes for thermal $\CN=4$ SYM on \AdS{4}. When $\rho_+ = 0$ we have our familiar friend, pure \AdS{5}, while $\rho_+ \neq 0$ provides the \SAdS{5} geometry written in coordinates where constant $\rho$ slices are (conformal to) copies of \AdS{4}. Since with $\rho_+ \neq 0$, the temporal Killing field has vanishing norm at $\rho =\rho_+$, regularity of the Euclidean geometry requires that one identify the Euclidean time coordinate with period $\beta = T_H^{-1}$. Here $T_H$ is the Hawking temperature of a \SAdS{d+1} black hole
\begin{equation}
T_H = \frac{d \, \rho_+^2 + (d-2)\, L_{d+1}^2}{4 \pi\, \rho_+ \, L_{d+1}^2} \ .
\label{bhtemp}
\end{equation}
Note that the \SAdS{d+1} solutions have a minimum Hawking temperature attained at $\rho_+ = \sqrt{\frac{d-2}{d}} L_{d+1}$, with $T_\text{min} = \frac{1}{2\pi\, L_{d+1}} \, \sqrt{d\,(d-2)}$.

Let us now return to the thermal $\CN =4$ SYM on \AdS{4}, where we impose the quotient boundary conditions preserving sixteen supercharges. From \req{bhtemp} we see that the \SAdS{5} solutions with $\rho_+ \neq 0$ can provide potential holographic duals only for $T \ge \frac{\sqrt{2}}{\pi\, \ell_4}$.\footnote{We henceforth use the geometry \req{warpads} to express the temperatures in terms of the field theory scale $\ell_4$.} For lower values of the temperature there is a unique geometry, viz., the thermal \AdS{5} geometry written in \AdS{4} foliation (together with the $\ZZ_2$ involution).

However, one still has to check which of these geometries has lower free energy. This can be determined for instance by computing the Euclidean action on the solutions. In fact, there is nothing new to compute as the results are easily obtained by recalling the physics of the Hawking-Page transition \cite{Hawking:1982dh,Witten:1998zw}.\footnote{Note that we are actually interested in quotients of \req{warpads} to describe duals of the field theories discussed in \sec{s:bcinvo}, but the effect of the quotient is just a rescaling of the free energy.} For the spacetimes of the general form \req{adsglobal},  one finds that the \SAdS{5} black hole solution has lower free energy only for $\rho_+ \ge L_{5}$.\footnote{In fact this statement is true in all dimensions; the black hole solution dominates only when its radius exceeds the AdS radius. It follows then that the Hawking-Page temperature for a $d$-dimensional CFT is simply $T_c =\frac{d-1}{2\pi} \, \frac{1}{\ell_{d}}$.} This implies that the Hawking-Page transition for strongly coupled $\CN =4$ SYM with quotient boundary conditions occurs at a critical temperature $T_c = \frac{3}{2\pi}\,\frac{1}{\ell_{4}}$.

From an explicit evaluation of the Euclidean action, one can argue that the free energy of the theory as ascertained from the holographic description behaves as follows \cite{Witten:1998zw}:
\begin{equation}
F =
\begin{cases}
& \CO(1) \ , \qquad T < T_c\\
& \CO(N^2) \ , \qquad T > T_c
\end{cases}
\label{}
\end{equation}	
One thus sees a sharp jump in the free energy in the strict large $N$  limit (which would be smoothed out into a cross-over at finite $N$). The transition at $T_c$ in the holographic description is expected to be the strong coupling analog of the large $N$ Hagedorn transition which we discussed in \sec{s:involweak}.\footnote{Both the strong coupling transition at $T_c$ and the weak coupling transition at \req{Hagquo} occur at precisely the same temperature as for the ${\cal N}=4$ SYM on an ${\bf S}^3$ with the same radius of curvature \cite{Sundborg:1999ue,Aharony:2003sx}. The relation between this theory and ours is clear in the holographic dual, but is not obvious directly in the field theory.}

In summary, for the boundary conditions involving quotient constructions with Neumann boundary conditions for a large subgroup $H$ of $G$ (specifically $\text{dim}({\mathfrak h}) \sim N$), we have argued that the holographic dual geometries are given by rewriting known asymptotically globally \AdS{5} solutions in a conformal frame where the boundary is \AdS{4}, combined with a $\ZZ_2$ action.  The phase structure for this class of theories exhibits qualitative similarities between the weak coupling and strong coupling results.

\section{Discussion and open questions}
\label{s:discuss}

In this paper we have investigated the dynamics of conformal field theories on \AdS{4}, with specific attention to ${\cal N} = 4$ SYM.  Due to the existence of a vast spectrum of boundary conditions admissible for the fields on the (timelike) boundary of \AdS{4}, we have seen that the physics of such field theories is quite rich, with non-trivial phase structure in certain cases. Part of our motivation for concentrating on the specific example of ${\cal N} =4$ SYM was due to the fact that we could discuss it at strong coupling both using S-duality and holographic methods.

Understanding how S-duality acts on the boundary conditions required us to understand its action on the
supersymmetric vacua of the theory on \AdS{4} (or equivalently on the Nahm data in the boundary
conditions). In some cases we also needed to understand
boundary degrees of freedom, living on the $2+1$ dimensional boundary of \AdS{4}, including \cite{Gaiotto:2008sa} boundary CFTs and/or charged matter (which can be realized in terms of additional D5 or NS5-branes for unitary gauge groups).

One of the main results of the analysis has been that the simplest class of boundary conditions which we call Dirichlet or Neumann, do not lend themselves readily to a holographic treatment. In the Dirichlet case, this was due to the presence of a large number of light excitations in the theory, which survive at strong coupling. The Neumann boundary conditions seem to be more tractable using holographic methods, but in order to determine if this is so, one needs to look for appropriate solutions of Type IIB supergravity describing the near-horizon limit of D3-branes ending on five-branes, which we postpone to future work. Note that \cite{DHoker:2007xy,DHoker:2007xz} claim to find all supergravity solutions with the appropriate symmetry, but it is not clear if and how the solutions we require are included in their classification.

Given the impasse for the basic boundary conditions permitted for the theory, we described how one can work with more general boundary conditions involving quotients that preserve supersymmetry \cite{Gaiotto:2008ak}. One can view these boundary conditions as Neumann boundary conditions for a subgroup $H$ of the gauge group $G$. We saw that provided $H$ was sufficiently large, it was possible to analyze these theories holographically in terms of  geometries which involve \AdS{4} foliations of \AdS{5}, together with a ${\mathbb Z}_2$ quotient which acts via a standard orientifold or orbifold action in string theory.  Such holographic duals have ${\cal O}(1)$ light degrees of freedom as one would expect from studying the weakly coupled CFT.

 In our analysis of ${\cal N} =4$ SYM, it should be borne in mind that we have restricted attention to boundary conditions that preserve sixteen supercharges. It would be interesting to determine whether there are examples of boundary conditions which are tractable at strong coupling despite preserving less supersymmetry. For example, there exist involutions that preserve less than 16 supercharges, and the dual geometries are again appropriate quotients of \AdS{5} $\times\ {\bf S}^5$.
 At the same time it would be interesting to determine if there are stable non-supersymmetric boundary conditions. Typically, breaking all supersymmetry tends to lead \cite{Berkooz:1998qp} (at least at finite $N$) to either perturbative instabilities (twisted sector tachyons in quotient constructions, violations of the Breitenlohner-Freedman bound, or higher trace relevant operators that break conformal invariance) or non-perturbative ones (brane nucleation or tunneling instabilities).  Another potential avenue for exploration in four dimensions is to consider the large class of ${\cal N} =1$ superconformal theories which arise from D3-branes probing various singularities. It would also be interesting to check if there is any supersymmetric index that can be computed for our superconformal theories on \AdS{4} (with SUSY-preserving boundary conditions). Such an index may be used to compare the theories at weak and strong coupling, or to learn about the strong coupling behaviour in the cases where it is not yet understood.

The boundary conditions we have chosen to examine for the gauge fields in \AdS{4} have been either Neumann or Dirichlet. However, as described in earlier work on boundary conditions for vector fields \cite{Witten:2003ya,Marolf:2006nd}, one can impose more general boundary conditions which relate the two fall-offs of the gauge field via a functional relation (similar to the multi-trace boundary conditions for scalar fields in the window around the BF bound, as discussed in \cite{Berkooz:2002ug,Witten:2001ua}). For a linear relation between the two fall-off coefficients, one can describe the boundary conditions in terms of D3-branes ending on $(p,q)$-fivebranes \cite{Gaiotto:2008sd}. For the Abelian theory this corresponds to addition of a boundary Chern-Simons term \cite{Witten:2003ya}. It would be interesting to examine the consequences of such boundary conditions for $d=4$ CFTs, and, in cases like ${\cal N}=4$ SYM, how they behave under S-duality (and its $SL(2,\ZZ)$ extension) and the AdS/CFT correspondence.

The study of CFTs on \AdS{} spacetime is by no means restricted to four spacetime dimensions. One can easily extend the analysis to lower or higher dimensions; our main reason for sticking to $d=4$ was due to the fact that, with the standard choice of norm and for the usual Yang-Mills kinetic term, gauge fields enjoy a rich set of boundary conditions only in this dimension. In higher dimensions, one may choose a different norm  that again allows a rich class of boundary conditions, but one expects this procedure to introduce ghosts \cite{Compere:2008us}. In three dimensions, one can have Chern-Simons gauge fields, and these are now known to play an important role in the construction of superconformal theories \cite{Gaiotto:2007qi,Aharony:2008ug,Aharony:2008gk}. In the special case of the CFTs related to M2-branes, there is a preliminary analysis of potential boundary conditions for such theories (once again on the half-space) \cite{Berman:2009kj,Chu:2009ms,Berman:2009xd,Chuetal}, which should allow one to study these three dimensional superconformal theories on the \AdS{3} geometry. However, this seems to include only the boundary conditions related to M2-branes ending on M5-branes; as above, we do not expect this case to lead to smooth holographic duals, while there should be alternative boundary conditions related to M2-branes on orbifolds/orientifolds, whose holographic duals are given by M theory on $(\text{AdS}_{4} \times {\bf S}^7)/\ZZ_2$. In addition, in M-theory one can also consider M5-branes ending on orbifolds, which should have a holographic dual in terms $(\text{AdS}_{7} \times {\bf S}^4)/\ZZ_2$ spacetimes; the challenge here is to understand the relevant boundary conditions directly in the field theory.

Finally, while our analysis has been restricted to the realm of conformal field theories, the \AdS{} spacetimes provide an interesting background also for theories with mass scales, especially confining gauge theories \cite{Callan:1989em,abty}. This is because they provide a geometric infra-red regulator for the theory. Understanding the consequences of the boundary conditions and dynamics of such confining theories on \AdS{} spacetimes promises to be an interesting avenue for exploration and we hope to return to this problem in the future.

\subsection*{Acknowledgments}
\label{s:acks}

It is a pleasure to thank Micha Berkooz, Veronika Hubeny, Andreas Karch, David Kutasov, Simon Ross, Nathan Seiberg, David Tong, and Shimon Yankielowicz for useful discussions.
MR would like to thank the Weizmann Institute of Science where this project was initiated for their wonderful hospitality. OA would like to thank Durham University and the LMS symposium on ``Non-perturbative techniques in field theory'' for hospitality during the course of this project. In addition MR would also like to thank the Amsterdam String Workshop, the University of  British Columbia and the Galileo Galilei Institute, Firenze for hospitality during the course of the project. The work of OA was supported in part by the Israel--U.S.~Binational Science Foundation, by a research center supported by the Israel Science Foundation (grant number 1665/10), by a grant (DIP H52) of the German Israel Project Cooperation, and by the Minerva foundation with funding from the Federal German Ministry for Education and Research. DM was supported in part by the US National Science Foundation under grants  PHY05-55669 and PHY08-55415 and by funds from the University of California.  MR is supported in part by an STFC Rolling grant.

\appendix

\section{Review of boundary conditions for fields with spin $s \le 1$}
\label{s:bcreview}
In this appendix we summarize the basic facts about boundary conditions for various matter fields in \AdS{d} spacetimes. As in \sec{s:qftads} we work in global coordinates and write down the possible boundary conditions for scalars, vectors and spinor fields. One obtains this information by considering the fall-off conditions and imposing appropriate normalizability conditions for the various fields.

\subsection{Scalar fields in \AdS{d}}
\label{s:scalarbc}

Consider a free classical scalar field in \AdS{d} with the metric \req{global2} and the action
\begin{equation}
\CS_\text{scalar} = \int d^dx \, \sqrt{-g} \, \left(\frac{1}{2}\, \nabla_\mu \phi \, \nabla^\mu \phi + \frac{1}{2}\, m^2 \, \phi^2  \right) .
\label{scact}
\end{equation}	
The boundary conditions which can be imposed on the scalar are well known from the early work of
\cite{Breitenlohner:1982jf} and have been discussed in the AdS/CFT context in
\cite{Klebanov:1999tb}. They depend only on the mass of the scalar and are insensitive to
 any couplings.\footnote{In particular, the boundary conditions are not affected by replacing the covariant derivative by a gauge covariant derivative; $\nabla_\mu \to\CD_\mu = \nabla_\mu - i A_\mu$.}

First of all,  one requires that the scalar mass lie above the so-called Breitenlohner-Freedman (BF) bound
\begin{equation}
m^2 \ge m^2_{BF} = -\frac{(d-1)^2}{4\, \ell_d^2} \ , \qquad \text{for} \; \text{AdS}_d \ .
\label{}
\end{equation}	
This ensures the stability of the scalar field theory in the background. From the classical equations of motion, one derives easily the possible asymptotic behaviours of the scalar:
\begin{equation}
\phi(r) \simeq r^{-\Delta\pm} \ , \qquad \Delta_\pm = \frac{d-1}{2} \pm \sqrt{\frac{(d-1)^2}{4} + m^2 \, \ell_d^2}.
\label{sfalloff}
\end{equation}	

It follows that the available boundary conditions for scalar fields in \AdS{d} are:
\begin{itemize}
\item If $m^2 > m_{BF}^2 + \ell_d^{-2}$ then only the mode that behaves as $r^{-\Delta_+}$ is normalizable with respect to the standard Klein-Gordon norm of the scalar field, so the other mode must be fixed. These are the usual boundary conditions imposed for scalar fields in AdS/CFT, where the scalar of mass $m$ is dual to an operator of dimension $\Delta = \Delta_+$ in the dual CFT.
\item If $m_{BF}^2 \le m^2 \le m_{BF}^2 + \ell_d^{-2}$ then both the modes $r^{-\Delta_\pm}$ are normalizable. In particular, this means that we can consider two particularly natural choices of boundary condition. Choosing $\phi \sim r^{-\Delta_+}$ is the conventional boundary condition. On the contrary, as  discussed in \cite{Klebanov:1999tb}, the choice $\phi \sim r^{-\Delta_-}$ (with the sub-leading ${\cal O}(r^{-\Delta_+})$ term fixed) leads to the scalar being dual to an operator of dimension $\Delta = \Delta_-$ in a putative dual CFT.  It is this choice that allows one to reach the unitarity bound on scalar operators for CFTs on \ESU{d-1}, since $\text{min}(\Delta_-) = \frac{d-1}{2}  -1 $.
\end{itemize}

Before we proceed to discuss other matter fields let us remark on one other scalar coupling of interest.  We will mainly consider CFTs on \AdS{d}, which generally  have scalar fields which couple to the background curvature.  Such a conformally coupled scalar field has an action:
\begin{equation}
\CS_\text{conformal scalar} = \int d^dx \, \sqrt{-g} \, \left(
 \frac{1}{2}\, \nabla_\mu \phi \, \nabla^\mu \phi   + \frac{d-2}{8\,(d-1)} \, R \, \phi^2 + \frac{1}{2}\, m^2 \, \phi^2\right).
\label{confkg}
\end{equation}	
Since the scalar curvature of \AdS{d} is constant
\begin{equation}
R = -\frac{d(d-1)}{\ell_d^2} \ ,
\label{}
\end{equation}	
one finds that a conformal coupling simply shifts the mass of the scalar field
\begin{equation}
m^2 \to m_c^2+ m^2 \ , \qquad m_c^2 \equiv-\frac{d(d-2)}{4\,\ell_d^2}.
\label{confmass}
\end{equation}	
The special case of a massless $m^2 = 0$ conformally coupled scalar field will be relevant for the study of CFTs on AdS spacetimes. For such fields, note that the conformal mass $m_c^2$ lies in the  middle of the interesting window where both boundary conditions are allowed; one has $\Delta_+ = \frac{d}{2}$ and $\Delta_- = \frac{d-2}{2}$.

Once we have the boundary conditions at hand, it is easy to ascertain the spectrum of a free scalar field. Exploiting the manifest $\R_t \times SO(d-1)$ isometry of the global \AdS{d} spacetime \req{global2} one can expand in modes
\begin{equation}
\phi  = e^{-i \, \omega \, t}\, {\bf Y}_{(k)}(\Omega)\, F(r),
\label{}
\end{equation}	
with ${\bf Y}_{(k)}(\Omega)$ ($k \ge 0$) being the spherical harmonics on $\Sp^{d-2}$. We recall that these scalar harmonics have eigenvalues
\begin{equation}
\nabla^2_{\Sp^{d-2}} \,{\bf Y}_{(k)}= -k\, (k+d-3) \, {\bf Y}_{(k)}\ , \qquad k = 0, 1, 2 , \cdots
\label{}
\end{equation}	
 Note that we are suppressing the analogs of the azimuthal quantum numbers and therefore have to remember that the spherical harmonics on $\Sp^{d-2}$ labeled by $k$ have a degeneracy $D_s(d-2,k)$ for\footnote{This degeneracy can be easily computed by realizing the scalar harmonics ${\bf Y}_{(k)}$ on $\Sp^{d-2}$ as projections of harmonic functions from $\R^{d-1}$. The latter are simply expressed in terms of symmetric, traceless,  homogeneous polynomials of degree $k$ which can be enumerated to give \req{shardeg}.}
\begin{equation}
D_s(d,k) = \;\; ^{d+k}\CC_{k} - \;\; ^{d-2+k}\CC_{k-2} =  \frac{d + 2\, k - 1}{d-1} \;\;\; ^{d+k-2}\CC_{k},
\label{shardeg}
\end{equation}	
where $^n\CC_k \equiv \frac{n!}{k! \,(n-k)!}$.
 It is well known that the Klein-Gordon equation arising from \req{scact} can be solved in terms of hypergeometric functions (see for instance \cite{Aharony:1999ti}). Demanding that the field be non-singular in the interior and normalizable, one immediately obtains a discrete spectrum (intuitively this follows because \AdS{d} acts as a confining box):
\begin{equation}
\omega\,\ell_d = 2 \, n + k + \Delta\ , \qquad \Delta = \Delta_\pm \ , \qquad n = 0, 1, 2, \cdots
\label{scalarspec}
\end{equation}	
and we have a total of $D_s(d-2,k)$ such states for a given harmonic at level $k$.

\subsection{Vector fields in  \AdS{d}}
\label{s:vectorbc}

Boundary conditions for massless vector fields propagating in asymptotically AdS spacetimes present a more involved story. In the early discussion of \cite{Breitenlohner:1982jf}, vector fields in \AdS{4} were discussed in the context of gauged supergravity theories, and it was realized in \cite{Witten:2003ya} that non-trivial boundary conditions are engendered in the context of \AdS{4}/CFT$_3$ duality.  The comprehensive analysis of the behaviour of Maxwell fields was carried out in \cite{Ishibashi:2004wx} which we briefly summarize below. These boundary conditions were used in  \cite{Marolf:2006nd} to obtain non-trivial bulk-boundary pairs in the context of AdS/CFT, generalizing the earlier analysis of  \cite{Breitenlohner:1982jf,Witten:2003ya}.

The basic idea behind the analysis of  \cite{Ishibashi:2004wx}  is to ask what are the boundary conditions that one should impose on the time-like $\scri^+$ of \AdS{d} for gauge fields propagating in the spacetime. One has the standard Maxwell action\footnote{We will explicitly discuss the boundary conditions for Abelian gauge fields. The generalization to the non-Abelian case is straightforward.}:
\begin{equation}
\CS_\text{gauge} = \frac{1}{e^2} \,\int d^dx \, \sqrt{-g} \, \left(
-\frac{1}{4}\, F_{\mu\nu} \, F^{\mu\nu} \right) \ , \qquad F = dA.
\label{maxact}
\end{equation}	
 The boundary conditions arise by examining the solutions to the Maxwell equations:
\begin{equation}
\nabla_\mu F^{\mu\nu}= 0 \ , \qquad \nabla_{[\mu}F_{\rho\sigma]} = 0.
\label{maxeom}
\end{equation}	

There are two basic issues to deal with for these equations: the gauge invariance and the fact that the gauge potential $A$ transforms as a vector field in \AdS{d}.  Once again using the manifest $SO(d-1) $ isometry of \req{global2}, the 1-form $A$ can be decomposed into
\begin{equation}
A = A^{\bf v} + A^{\bf s},
\label{adecomp}
\end{equation}	
where $A^{\bf v}$ is the vector of $SO(d-1)$ and $A^{\bf s}$ is a scalar with respect to $SO(d-1)$. In terms of a harmonic decomposition, $A^{\bf v}$ can be expanded in vector spherical harmonics ${\bf V}^i_{(k)}$ and $A^{\bf s}$ in terms of scalar harmonics ${\bf Y}_{(k)}$ and their derivatives.

To facilitate the discussion it is useful to introduce a basis of one-forms on \AdS{d}; we'll take $dx^a$ to be one-forms in the two dimensional space spanned by $(t,r)$, and $d\Omega^i$ to be the angular one-forms.  Then it is easy to see that the harmonic decomposition takes the form
\begin{equation}
A^{\bf v} = \sum_{k} \; \Phi^{\bf v}_{(k)}(t,r)\, \, {\bf V}_{i \, (k)}(\Omega) \, d\Omega^i\,,
\label{av}
\end{equation}	
and
\begin{equation}
A^{\bf s} =  \sum_{k} \; \left({\mathbf A}_{a\,(k)}(t,r)\,  {\bf Y}_{(k)}(\Omega)  \, dx^a + {\bf a}_{(k)}(t,r)\, \nabla_i {\bf Y}_{(k)}(\Omega)  \, d\Omega^i\right).
\label{as}
\end{equation}	
The vector harmonics ${\bf V}_{i\,(k)}$ on $\Sp^{d-2}$ are again labeled by a single quantum number $k$, in terms of which their  eigenvalues are determined to be
\begin{equation}
\nabla^2_{\Sp^{d-2}} {\bf V}_{i \, (k)} = -\left(k\,(k+d-3) -1 \right) {\bf V}_{i \, (k)}\ , \qquad
\nabla^i {\bf V}_{i \, (k)} =0 \ ,\qquad k = 1, 2 \, \cdots.
\label{}
\end{equation}	
Note that there are no vector harmonics for $k=0$.  As we have suppressed the other quantum numbers labeling these harmonics, we have to keep track of their degeneracy, $D_v(d-2,k)$ where\footnote{As for the scalar harmonics, one can obtain vector harmonics on $\Sp^d$  by projecting vectors in $R^{d+1}$ onto the tangent space of $\Sp^d$. The degeneracy computation can be determined by removing the modes which are scalars and derivatives thereof from the $SO(d+1)$ point of view, thereby restricting to just divergence free vectors on $\Sp^{d}$. }
\begin{equation}
D_v(d,k) = (d+1)\, \left( ^{d+k}\CC_{k} - \; ^{d-2+k}\CC_{k-2} \right) - \;^{d+k+1}\CC_{k+1} - \; ^{d+k-3}\CC_{k-3}.
\label{}
\end{equation}	

The vector modes $A^{\bf v}$ are insensitive to gauge transformations. This makes it easy to read off  the boundary conditions for them once we know their fall-off conditions, which are
\begin{equation}
\Phi^{\bf v}_{(k)} = \alpha^{\bf v}_{(k)} + \beta^{\bf v}_{(k)} \, r^{3-d} \ , \qquad d \neq 3.
\label{vecmodefo}
\end{equation}	
This follows by realizing that the fields $r^{-\frac{d-4}{2}}\, \Phi^{\bf v}_{(k)}(r,t)$ effectively behave as massive scalars in an effective \AdS{2} geometry (the coset \AdS{d}$/\Sp^{d-2}$) with mass $m^2_v =  -\frac{d-2}{\ell_d^2}$. Note that there are actually $d-3$ scalars here corresponding to the different polarizations of the spherical harmonic, which should be kept in mind while counting the number of degrees of freedom.

On the other hand the scalar part $A^{\bf s}$ is affected by the underlying gauge invariance. This can in fact be fixed \cite{Ishibashi:2004wx} by noting that the source-free equations of motion set ${\bf A}_{a\, (k)} = {\bf a}_{(k)} = 0$ for $k=0$, and that for $k \ge 1$ we may trade the unique gauge-invariant combination of ${\bf A}_{a\, (k)}$ and ${\bf a}_{(k)}$ for another scalar field $\Phi^{\bf s}_{(k)}$, which determines the mixed components of the gauge field strength:
\begin{equation}
\epsilon_{ab}\,\left(\nabla^b {\bf a}_{(k)}  + \sqrt{k(k+d-3)} \, {\bf A}^b_{\, (k)} \right) =\frac{1}{r^{d-4}}\,  \nabla_a \Phi^{\bf s}_{(k)}\,.
\label{Phisdef}
\end{equation}	
The fall-off conditions for $\Phi^{\bf s}_{(k)}$ can then be determined from the equations of motion to be
\begin{eqnarray}
\Phi^{\bf s}_{(k)} &=& \alpha^{\bf s}_{(k)} \, r^{d-5} + \beta^{\bf s}_{(k)} \, r^{0} \ , \qquad d \neq 5, \nonumber \\
&=& \alpha^{\bf s}_{(k)} \, \log (r) + \beta^{\bf s}_{(k)} \, r^{0} \ , \qquad d = 5.
\label{scmodefo}
\end{eqnarray}	
One can again encode this information by realizing that the field $r^\frac{d-4}{2}\, \Phi^{\bf s}_{(k)}(r,t)$ behaves as a scalar field of mass $
m^2_s  = - \frac{2\, (d-3)}{\ell_d^2} $ in the effective \AdS{2} geometry.

Intuitively, the scalars $\Phi^{\bf v}$ and $\Phi^{\bf s}$ correspond to the magnetic and electric parts of the gauge field. Rather than elaborate on their properties in various dimensions, we present the relevant details regarding the boundary conditions for the cases we will be interested in and refer the reader to \cite{Ishibashi:2004wx} for the general story. The comprehensive description of the allowed boundary conditions in general dimensions is summarized in \cite{Marolf:2006nd}.

This somewhat abstract discussion can be translated into the $A_r =0$ gauge used in \sec{s:qftads}. In order to do so, it is useful to convert the information in $\Phi^{\bf s}_{(k)}$ and $\Phi^{\bf v}_{(k)}$ into the gauge potential itself. From \cite{Marolf:2006nd} we learn that
\begin{eqnarray}
A_i &=& D_i \, \lambda + \alpha_i(\Omega^i,t) r^0 + \beta_i(\Omega^i,t) \,r^{2-d} \nonumber \\
A_t &=& \partial_t\, \lambda + c_s(d)\,\alpha_s(\Omega^i,t) + {\cal O}(r^{2-d}) +{\cal O}(r^{-2})\nonumber \\
A_r &=& \partial_r \, \lambda + \partial_t\alpha_s(\Omega^i,t) r^{-3} + \partial_t\beta_s(\Omega^i,t) \,r^{1-d}
\label{Agensol}
\end{eqnarray}	
where $c_s(d) = d-5$ for $d\neq 5$ and is set to unity for $d=5$, and $\lambda(r,t,\Omega^i)$ is a scalar gauge function.  To write the expressions
we have taken linear combination of the harmonics to define functions $\alpha_{i,s}(\Omega^i,t)$ and  $\beta_{i,s}(\Omega^i,t)$ of the boundary coordinates via
\begin{equation}
f_i(\Omega^i,t) = \sum_{k} \, f_{(k)}^{\bf v} \, {\bf V}_{i\,(k)} \ , \qquad f_s(\Omega^i,t) = \sum_{k} \, f_{(k)}^{\bf s} \, {\bf Y}_{(k)}\,.
\label{}
\end{equation}	
Solving the last equation of \req{Agensol}  to set $A_r =0$ we can find $\lambda(r,t,\Omega^i)$. Plugging the resulting expression into the boundary components of the gauge field leads to \req{Afalloff}.

\paragraph{Gauge fields in \AdS{4}:} A Maxwell field in four dimensions has two propagating degrees of freedom. In terms of the decomposition \req{adecomp} one can view these as being the magnetic scalar arising from the vector mode $A^{\bf v}$, and the electric scalar $\Phi^{\bf s}$. Furthermore, in \AdS{4} one expects (for a free Maxwell theory) these scalars to be equivalent because of classical electric-magnetic duality of  \req{maxact}. This is indeed true, as in $d=4$ the scalars $\Phi^{\bf v}$ and $\Phi^{\bf s}$ have the same mass. Moreover, this value is precisely that of a conformally coupled scalar field in \AdS{4}, which also follows from the classical conformal invariance of \req{maxact}.

Since the two physical degrees of freedom of the gauge field in \AdS{4} obey the conformal scalar wave equation,  it follows that both these modes can be allowed to have either of the two natural boundary conditions discussed above. It is physically more intuitive however to phrase the boundary conditions in terms of the electric and magnetic components of the gauge field. While one can think of the electric component as related to $\Phi^{\bf s}$, the magnetic component is really related to $\Phi^{\bf v}$ via electric-magnetic duality.

The `standard boundary conditions', which we will refer to as {\em Dirichet bc} for the gauge field, correspond to fixing the components of $F$ tangential to the boundary of \AdS{4} (the boundary value of the gauge field is declared to be a non-fluctuating mode). This translates to using a Dirichlet boundary condition for the electric scalar, but a Neumann boundary condition for the magnetic scalar. These boundary conditions allow for electrically charged states in \AdS{4}, which are charged under a global $U(1)$ symmetry that the theory has (corresponding
to constant gauge transformations, which do not vanish at the boundary so they act non-trivially
on the Hilbert space). In the AdS/CFT correspondence, the operator dual to the bulk field $A_{\mu}$ for these boundary conditions is a conserved current
$J^{i}$ on ESU$_3$, $\partial_{i} J^{i} = 0$; but note that the theory has a global symmetry
even without using the AdS/CFT correspondence. In this case the magnetic scalar has Neumann boundary conditions, so one
has a Gauss' law constraint forbidding magnetically charged  states in the theory.

On the other hand, one can impose `modified boundary conditions' which we will refer to as {\em Neumann bc}, as described in \cite{Witten:2003ya,Marolf:2006nd}.  These correspond to imposing Dirichlet boundary conditions on the magnetic scalar and Neumann boundary conditions on the electric scalar. This therefore involves fixing the components of the bulk gauge field which have one leg in the radial direction of \AdS{4}. These modified boundary conditions are a bit more interesting: the gauge field $A_{\mu}$ is allowed to fluctuate on the boundary, so here we end up having dynamical gauge symmetry on the $\partial$\AdS{4} $=$ \ESU{3}, and as a result disallow states carrying non-zero total charge. One can however consider magnetically charged states in the theory, since now there is no Gauss' law constraint for the magnetic gauge field. In the
context of the AdS/CFT correspondence, the dual CFT$_3$ now has a dynamical gauge field
(though with no three dimensional kinetic term) instead of a conserved global current, and
the lowest-dimensional gauge-invariant operator is its field strength. In the Abelian case, electric-magnetic duality relates this field strength to a global symmetry current by $J = *_3 F$.

The spectrum of the gauge fields in \AdS{4} is captured completely by the conformal scalars $\Phi^{\bf s, v}$ introduced earlier. This implies that the energy levels are a subset of
\begin{equation}
\omega\,\ell_d = 2 n + l + \Delta_\pm \ , \qquad \Delta_\pm =1 ,2 \ , \qquad n = 0, 1, 2, \cdots
\label{vecspec}
\end{equation}	
The standard (Dirichlet) boundary condition treats $\Phi^{\bf s}$ as a dimension 1 operator and $\Phi^{\bf v}$  as a dimension 2 operator. On the other hand the modified (Neumann) boundary conditions entail treating $\Phi^{\bf s}$ as a dimension 2 operator and $\Phi^{\bf v}$  as a dimension 1 operator. Since the scalars are isomorphic in the bulk, the spectral content of the theory is identical in both cases. The only issue to worry about is the one described earlier of having to impose Gauss' law for electrically charged or magnetically charged states in the theory.

Note that the boundary condition for the gauge field affects also other fields; for instance, if we have
a Neumann boundary condition for the gauge field, we cannot introduce sources for any electrically charged
fields in the bulk, since such sources would not be gauge-invariant. This corresponds in a putative dual CFT
to the fact that only correlation functions of gauge-invariant operators are well-defined.

\subsection{Fermions in \AdS{d}}
\label{s:fermionbc}

Having discussed scalars and vectors in \AdS{d} we now turn our attention to fermionic fields. The boundary conditions for fermions were worked out first in \cite{Breitenlohner:1982jf}. An early discussion in the context of the AdS/CFT correspondence can be found in \cite{Henningson:1998cd,Mueck:1998iz} while \cite{Amsel:2008iz, Iqbal:2009fd} provide a more recent perspective.

As before it suffices to discuss the free fermion action:
\begin{equation}
\CS_{\text{fermion}} =\int\, d^{d} x \, \sqrt{-g} \, i \left({\bar \psi} \, \Gamma^\mu \nabla_\mu \psi - m \, {\bar \psi} \psi\right),
\label{feract}
\end{equation}	
where the spinor covariant derivative is as usual expressed in terms of the spin-connection
\begin{equation}
\nabla_\mu = \partial_\mu + \frac{1}{4}\, \omega_{\mu pq} \, \gamma^{pq}\,.
\label{diracop}
\end{equation}	
Note that we are now using lowercase latin letters ($p,q,\cdots \in \{t,r,i\}$) to denote the local tangent space index. To be specific, for the metric \req{global2}, one can take a basis of veilbeins
\begin{equation}
e_t = \sqrt{f(r)} \, dt  \ , \quad e_r = \frac{1}{\sqrt{f(r)}} \, dr \ , \quad
e_i = r\, \hat{e}_i  \ , \qquad {\rm where} \;\; f(r) =1+\frac{r^2}{\ell_d^2}\,,
\label{}
\end{equation}	
with $\hat{e}_i$ a standard basis of one-forms on $\Sp^{d-2}$. One then has $\Gamma^\mu = e^{\mu}_{\;p}\, \gamma^p$.

One then finds that the spin connections are
\begin{equation}
\omega_{tr} = \frac{f'(r)}{2}\, dt \ , \quad \omega_{ri} = -\sqrt{f(r)} \, \hat{e}_i \ , \quad \omega_{ij} = \hat{\omega}_{ij}\,,
\label{}
\end{equation}	
where once again $\hat{\omega}_{ij}$ is the spin-connection on $\Sp^{d-2}$. The Dirac operator can then be shown to take the form:
\begin{equation}
\Gamma^\mu\,\nabla_\mu = \sqrt{f} \, \gamma^r \, \partial_r  - \frac{1}{\sqrt{f}} \, \gamma^t\,\partial_t +\left(\frac{d-2}{2}\, \frac{\sqrt{f}}{r} +\frac{f'}{4\,f}\right) \gamma^r+\frac{1}{r}\, \Gamma^i\,\hat{\nabla}_i\,.
\label{}
\end{equation}	
One can solve this equation as in \cite{Breitenlohner:1982jf} by using a spinor harmonic decomposition.

However, in order to ascertain the boundary conditions for fermions, it actually suffices to work in the Poincar\'e patch, and then transcribe the result to the global coordinates. In the Poincar\'e coordinates one finds that the Dirac operator takes the form:
\begin{equation}
\Gamma^\mu\,\nabla_\mu = r \, \gamma^r \, \partial_r  + i \, \frac{1}{r}\,\gamma^\mu\,p_\mu + \frac{d-1}{2}\,r.
\label{}
\end{equation}	
From the free Dirac equation, one can then easily ascertain that \cite{Iqbal:2009fd}
\begin{equation}
\psi_\pm(r) \to a_\pm(x) \, r^{-\frac{d}{2} \pm m\ell_d} + b_\pm(x)\, r^{-\frac{d}{2} \mp m\ell_d -1}  \  \qquad \text{as} \;\; r \to \infty \ ,
\label{psifo}
\end{equation}	
with
\begin{equation}
\psi_\pm = \Gamma_\pm \, \psi \ , \qquad \Gamma_\pm = \frac{1}{2}\,\left(1\pm \Gamma^r\right).
\label{psipm}
\end{equation}	
These fall-off conditions suffice to determine the boundary conditions for the fermions as we now explain.

For fermions there are two important issues to keep in mind whilst prescribing boundary conditions. One is that the Dirac equation being first order, Dirichlet boundary conditions can be quite constraining.  Secondly, the spinor representations in the bulk \AdS{d} and boundary \ESU{d-1} could have different number of components. As described in \cite{Amsel:2008iz,Iqbal:2009fd} these two facts conspire to produce a sensible set of boundary conditions for fermions, once one realizes that $\psi_\pm$ introduced above are conjugate variables.

The standard boundary conditions for fermions then correspond to fixing $a_+(x)$ on the boundary \ESU{d-1} for $m >0$ (correspondingly $a_-(x) $ for $m<0$). From these fall-off conditions one learns that a Dirac fermion of mass $m$ can be associated with an operator of conformal dimension:
\begin{equation}
\Delta = \frac{d-1}{2} + |m\,\ell_d|.
\label{ferdim}
\end{equation}	
Since the spinor representations are dimension specific it is useful to record the results for odd and even spacetime dimensions in turn:
\begin{itemize}
\item  In odd spacetime dimensions,  a Dirac spinor $\psi$ propagating in the bulk of \AdS{d}, can be mapped to a chiral spinor on the boundary \ESU{d-1}. Moreover, the chirality of the spinor is correlated with the sign of the mass term in \req{feract}; positive chirality implies we should impose Dirichlet boundary conditions on $\psi_+$ and hence $m>0$.
\item In even spacetime dimensions, on the other hand, the Dirac spinor $\psi$ in \AdS{d}, can be related to a Dirac spinor on \ESU{d-1}.
\end{itemize}

Note that \req{ferdim} suggests that fermionic fields in \AdS{d} are well gapped from the unitarity bound for fermionic operators in a CFT. Recall that the unitarity bound for fermions on \ESU{d-1} is actually $\frac{d-2}{2}$, in contrast to that of bosonic operators dimensions  which as described in \sec{s:scalarbc} is $\frac{d-1}{2} -1$. As in the scalar case, it turns out that one can indeed saturate the unitarity  bound, by realizing that in the regime of masses $0\le m\ell_d < \frac{1}{2}$, both fall-offs in \req{psifo} are normalizable. This in particular implies that in this range we can take $\Delta = \frac{d-1}{2} - |m\ell_d|$.

\section{Computing the Hagedorn temperature for the quotients}
\label{s:NeuHag}

We wish to compute the Hagedorn transition temperature for theories obtained by the various involutions of $SU(N)$. As described in \sec{s:involweak} the basic quantity of interest is the partition sum of the theory, which is given for a general gauge theory with  Neumann boundary conditions preserving $H \subset G$ in \req{HNeupf}. This formula needs to be evaluated in the various cases of interest. For simplicity we will do the computation for $G=U(N)$ rather than $SU(N)$; the two are equivalent in the large $N$ limit.

To do so let us first record the Haar measure on the classical groups in terms of the eigenvalues. For $U(N)$ one has eigenvalues which are pure phases $e^{i\, \theta_i}$, while for $USp(2N)$ and $SO(2N)$ there are $2N$ eigenvalues, which are pairs of conjugate phases, $\{e^{i\, \theta_i}, e^{-i\, \theta_i}\}$. For $SO(2N+1)$ one has an additional eigenvalue which is unity.  In terms of these eigenvalues, the Haar measure is given as (dropping irrelevant numerical pre-factors):
\begin{eqnarray}
U(N):&& \prod_{i=1}^N \, [d\theta_i] \, \prod_{1\leq i<j\leq N} \sin^2\left(\frac{\theta_i-\theta_j}{2} \right),  \nonumber \\
USp(2N): && \prod_{i=1}^N \, [d\theta_i] \, \prod_{1\leq i<j\leq N} ( \cos(\theta_i )- \cos (\theta_j))^2 \, \prod_{m=1}^N \, \sin^2(\theta_m),  \nonumber \\
SO(2N): && \prod_{i=1}^N \, [d\theta_i] \, \prod_{1\leq i<j\leq N} ( \cos(\theta_i) - \cos (\theta_j))^2 , \nonumber \\
SO(2N+1):&& \prod_{i=1}^N \, [d\theta_i] \, \prod_{1\leq i<j\leq N} ( \cos(\theta_i) - \cos (\theta_j))^2 \, \prod_{m=1}^N \, \sin^2 \left(\frac{\theta_m}{2}\right).
\label{Haarm}
\end{eqnarray}	

In addition we need the formulae for the characters of various representations. These are given again in terms of the eigenvalues above: denoting the irreducible representations of the classical groups by Young tableaux $(n_1, n_2\, \cdots n_N)$ with $n_1 \ge n_2 \ge \cdots \ge n_N \ge 0$ one has
\begin{eqnarray}
U(N): && \chi_{(n_1, n_2\, \cdots n_N)} (U)  =\frac{\text{det} \left[e^{i\,\theta_i \,(n_j + N -j)} \right]}{\text{det} \left[e^{i\,\theta_i \,(N -j)} \right]} \nonumber \\
USp(2N): && \chi_{(n_1, n_2\, \cdots n_N)} (U)  =\frac{\text{det} \left[e^{i\,\theta_i \,(n_j + N -j+1)} -e^{-i\,\theta_i \,(n_j + N -j+1)}\right]}{\text{det} \left[e^{i\,\theta_i \,(N +1-j)}- e^{-i\,\theta_i \,(N +1-j)}  \right]} \nonumber \\
SO(2N+1): && \chi_{(n_1, n_2\, \cdots n_N)} (U)  =\frac{\text{det} \left[e^{i\,\theta_i \,(n_j + N -j+\frac{1}{2})}-e^{-i\,\theta_i \,(n_j + N -j+\frac{1}{2})} \right]}{\text{det} \left[e^{i\,\theta_i \,(N+\frac{1}{2} -j)}-e^{-i\,\theta_i \,(N+\frac{1}{2} -j)} \right]} \nonumber \\
SO(2N): && \chi_{(n_1, n_2\, \cdots n_N)} (U)  = \frac{1}{2} \left(\frac{\text{det} \left[e^{i\,\theta_i \,(n_j + N -j)} + e^{-i\,\theta_i \,(n_j + N -j)} -\delta_{jN}\,\delta_{n_{N} 0}  \right]}{\text{det} \left[e^{i\,\theta_i \,(N -j)} +e^{-i\,\theta_i \,(N-j)} -\delta_{jN}\right]}    \right. \nonumber \\
&&\left. \qquad \qquad \qquad \qquad \qquad \pm\frac{\text{det} \left[e^{i\,\theta_i \,(n_j + N -j)}-e^{-i\,\theta_i \,(n_j + N -j)} \right]}
{\text{det} \left[e^{i\,\theta_i \,( -j)} + e^{-i\,\theta_i \,(N -j)} -\delta_{jN}\right]}\right)
\label{characters}
\end{eqnarray}	

For the cases of interest in \sec{s:bcquotients} the gauge groups and representations involved are given in  \req{branching}. Expressing this in terms of the Young tableaux data we have:
\begin{eqnarray}
U(p) \times U(q) : && \left(\big\{\yng(1), \overline{\yng(1)} \big\}, \{\bullet, \bullet\} \right) \oplus \left(\{\bullet, \bullet\} , \big\{\yng(1), \overline{\yng(1)} \big\} \right) \nonumber \\
&& \qquad
\oplus \left(\big\{\yng(1), \bullet \big\}, \big\{\bullet,\overline{\yng(1)}\big\} \right) \oplus \left(\big\{\bullet,\overline{\yng(1)}\big\}, \big\{\yng(1), \bullet \big\}\right) \nonumber
 \\
&& \nonumber \\
SO(N): && \yng(1,1) \oplus \yng(2) \nonumber \\
USp(N): &&  \yng(2) \oplus \yng(1,1)
\label{branchingn}
\end{eqnarray}	

The characters for the above representations can be easily obtained from \req{characters}. In particular, for the orthogonal and symplectic group representations occurring in \req{branchingn}\footnote{For simplicity, we restrict henceforth to odd $N$ for the orthogonal case. The corresponding formulae for even orthogonal groups can be derived straightforwardly.}
\begin{eqnarray}
\chi^{SO(N)}_{\tiny \yng(1,1)} + \chi^{SO(N)}_{\tiny \yng(2)} &=& \left(2\,\sum_i^{\frac{N-1}{2}}\, \cos(\theta_i) \right)\, \left(  2\,\sum_j^{\frac{N-1}{2}}\, \cos(\theta_j) +1\right) -1, \\
\chi^{USp(N)}_{\tiny \yng(2)} + \chi^{USp(N)}_{\tiny \yng(1,1)} &=& \left(2\,\sum_i^\frac{N}{2}\, \cos(\theta_i) -1 \right)\, \left(  2\,\sum_j^\frac{N}{2}\, \cos(\theta_j) +1\right).
\label{}
\end{eqnarray}	

With this data it is simple to write the integral \req{HNeupf} for the various cases. For simplicity denoting $z_B(x^m) + (-1)^{m+1} \, z_F(x^m) = \xi_m(x)$, one has generically an expression of the form:
\begin{equation}
Z(x) = \int_{-\pi}^{\pi} \, \prod_i\, [d\theta_i] \, e^{-{\cal V}(\theta_i, x) - {\cal V}_0}\,,
\label{}
\end{equation}	
with ${\cal V}_0$ an irrelevant constant.  The function of the eigenvalues in the exponential is given by:
\begin{eqnarray}
\CV^{U(p)\times U(q)}(\theta_i, {\hat \theta}_j, x) &=&\sum_{m=1}^\infty \,\frac{1}{m} \left(\left[1-\xi_m(x)\right]\, \left( \sum_{i\neq j=1}^p\,  \cos (m(\theta_i-\theta_j)) +  \sum_{i\neq j=1}^q \cos (m({\hat \theta}_i-{\hat \theta}_j)) \right) \right.\nonumber \\
&& \qquad \left. - \, 2 \, \xi_m(x) \,  \sum_{i=1}^p\, \sum_{j=1}^q  \cos (m(\theta_i-{\hat \theta}_j)) \right)
\label{}
\end{eqnarray}	
\begin{eqnarray}
\CV^{SO(N)}(\theta_i, x) &=& \sum_{m=1}^\infty \,\frac{1}{m}\left(2\,  [1-2\,\xi_m(x)] \, \sum_{i\neq j=1}^\frac{N-1}{2} \, \cos (m\, \theta_i) \, \cos (m \,\theta_j)  -\, (N-3) \, \xi_m(x)\right. \nonumber \\
&&   \left. \qquad+ 2\, [1-\xi_m(x) ] \,  \sum_{i=1}^\frac{N-1}{2} \, \big(\cos(2 m\, \theta_i) + \cos(m \,\theta_i) \big)\right)
\label{}
\end{eqnarray}	
\begin{eqnarray}
\CV^{USp(N)}(\theta_i, x) &=& \sum_{m=1}^\infty \,\frac{1}{m}\left( [1-2\,\xi_m(x)] \, \sum_{i\neq j=1}^\frac{N}{2} \, \cos (m\, \theta_i) \, \cos (m \,\theta_j)  \right. \nonumber \\
&& \qquad  \qquad \left. -\,(N-1) \, \xi_m(x) +2\, [1-\xi_m(x) ]  \, \sum_{i=1}^\frac{N}{2} \, \cos(2 m\,\theta_i) \right)
\label{}
\end{eqnarray}	

In the large $N$ limit one can convert the sums over the eigenvalues to integrals, and obtain an effective action for the eigenvalue distribution. Intuitively one expects that the repulsion between the eigenvalues arising from the measure \req{Haarm} will dominate at low temperatures leading to a uniform eigenvalue distribution. At sufficiently large temperatures it is possible that the attractive interaction arising from the thermal contribution leads to a gapped distribution. Introducing the eigenvalue distribution $\rho(\theta)$ one can write an effective action for the cases of interest in terms of the Fourier modes $\rho_n = \int_{-\pi}^\pi \, d\theta \, \cos(n\theta) \rho(\theta) $:
\begin{equation}
S_\text{eff}^{U(p)\times U(q)}\left[\rho(\theta),{\hat \rho}({\hat \theta}) \right]=\sum_{n=1}^\infty \, \frac{1}{n}\, \left( p^2 \, \rho_n^2 + q^2 \,{\hat \rho}_n^2- \big| p \,\rho_{n} + q\, {\hat \rho}_{n}\big|^2 \, \xi_n(x) \right),
\label{seffupuq}
\end{equation}	
\begin{equation}
S_\text{eff}^{SO(N)}\left[\rho(\theta)\right]=  \frac{(N-1)^2}{4}\, \sum_{n=1}^\infty \, \frac{1}{n}\, \left( [1-2 \,\xi_n(x)] \, \rho_n^2 +\frac{4}{N-1} \, [1-\xi_n(x)] \, (\rho_{2n} +\rho_n) - \frac{4\,(N-3)}{(N-1)^2} \,\xi_n(x) \right),
\label{}
\end{equation}	
\begin{equation}
S_\text{eff}^{USp(N)}\left[\rho(\theta) \right]= \frac{N^2}{4}\, \sum_{n=1}^\infty \, \frac{1}{n}\, \left( [1-2 \,\xi_n(x)] \, \rho_n^2 +\frac{4}{N} \, [1-\xi_n(x)] \, \rho_{2n} - \frac{4(N-1)}{N^2} \,\xi_n(x) \right).
\label{}
\end{equation}	
At large $N$ (and in the case of the unitary quiver $p,q \sim N$) we find that the uniform distribution becomes unstable when the Fourier mode $\rho_1$ goes tachyonic. This occurs in all cases at:
\begin{equation}
1- 2\, \xi_1 (x) = 0 \; \thus \;\; z_B(x) + z_F(x) = \frac{1}{2},
\label{}
\end{equation}	
as quoted in \req{Hagquo}. Note also that from the unitary quiver one can also obtain the standard result for a single unitary gauge group with matter in the adjoint. Setting $q =0$ in (\ref{seffupuq}) one obtains the expression for the Hagedorn transition given in \req{weakHagA}.


\begin{thebibliography}{10}

\bibitem{Breitenlohner:1982jf}
P.~Breitenlohner and D.~Z. Freedman, ``{Stability in Gauged Extended
  Supergravity},''
\href{http://dx.doi.org/10.1016/0003-4916(82)90116-6}{{\em Ann. Phys.} {\bf
  144} (1982)  249}.

\bibitem{Callan:1989em}
C.~G. Callan, Jr. and F.~Wilczek, ``{Infrared Behaviour at Negative
  Curvature},''
\href{http://dx.doi.org/10.1016/0550-3213(90)90451-I}{{\em Nucl. Phys.} {\bf
  B340} (1990)  366--386}.

\bibitem{Klebanov:1999tb}
I.~R. Klebanov and E.~Witten, ``{AdS/CFT correspondence and symmetry
  breaking},'' \href{http://dx.doi.org/10.1016/S0550-3213(99)00387-9}{{\em
  Nucl. Phys.} {\bf B556} (1999)  89--114},
\href{http://arxiv.org/abs/hep-th/9905104}{{\tt arXiv:hep-th/9905104}}.

\bibitem{Witten:2003ya}
E.~Witten, ``{SL(2,Z) action on three-dimensional conformal field theories with
  Abelian symmetry},''
\href{http://arxiv.org/abs/hep-th/0307041}{{\tt arXiv:hep-th/0307041}}.

\bibitem{Marolf:2006nd}
D.~Marolf and S.~F. Ross, ``{Boundary conditions and new dualities: Vector
  fields in AdS/CFT},'' {\em JHEP} {\bf 11} (2006)  085,
\href{http://arxiv.org/abs/hep-th/0606113}{{\tt arXiv:hep-th/0606113}}.

\bibitem{Ishibashi:2004wx}
A.~Ishibashi and R.~M. Wald, ``{Dynamics in non-globally hyperbolic static
  spacetimes. III: Anti-de Sitter spacetime},''
  \href{http://dx.doi.org/10.1088/0264-9381/21/12/012}{{\em Class. Quant.
  Grav.} {\bf 21} (2004)  2981--3014},
\href{http://arxiv.org/abs/hep-th/0402184}{{\tt arXiv:hep-th/0402184}}.

\bibitem{Karch:2000ct}
A.~Karch and L.~Randall, ``{Locally localized gravity},'' {\em JHEP} {\bf 05}
  (2001)  008,
\href{http://arxiv.org/abs/hep-th/0011156}{{\tt arXiv:hep-th/0011156}}.

\bibitem{Hubeny:2009rc}
V.~E. Hubeny, D.~Marolf, and M.~Rangamani, ``{Hawking radiation from AdS black
  holes},'' \href{http://dx.doi.org/10.1088/0264-9381/27/9/095018}{{\em
  Class.Quant.Grav.} {\bf 27} (2010)  095018},
  \href{http://arxiv.org/abs/arXiv:0911.4144}{{\tt arXiv:0911.4144
  [hep-th]}}.

\bibitem{Gaiotto:2008sa}
D.~Gaiotto and E.~Witten, ``{Supersymmetric Boundary Conditions in N=4 Super
  Yang-Mills Theory},'' \href{http://arxiv.org/abs/arXiv:0804.2902}{{\tt
  arXiv:0804.2902 [hep-th]}}.

\bibitem{Gaiotto:2008ak}
D.~Gaiotto and E.~Witten, ``{S-Duality of Boundary Conditions In N=4 Super
  Yang-Mills Theory},'' \href{http://arxiv.org/abs/arXiv:0807.3720}{{\tt
  arXiv:0807.3720 [hep-th]}}.

\bibitem{Compere:2008us}
G.~Compere and D.~Marolf, ``{Setting the boundary free in AdS/CFT},''
  \href{http://dx.doi.org/10.1088/0264-9381/25/19/195014}{{\em Class. Quant.
  Grav.} {\bf 25} (2008)  195014},
\href{http://arxiv.org/abs/0805.1902}{{\tt arXiv:0805.1902 [hep-th]}}.

\bibitem{Amsel:2009rr}
A.~J. Amsel and G.~Compere, ``{Supergravity at the boundary of AdS
  supergravity},'' \href{http://dx.doi.org/10.1103/PhysRevD.79.085006}{{\em
  Phys. Rev.} {\bf D79} (2009)  085006},
\href{http://arxiv.org/abs/0901.3609}{{\tt arXiv:0901.3609 [hep-th]}}.

\bibitem{Leigh:2003ez}
R.~G. Leigh and A.~C. Petkou, ``{SL(2,Z) action on three-dimensional CFTs and
  holography},'' {\em JHEP} {\bf 0312} (2003)  020,
  \href{http://arxiv.org/abs/hep-th/0309177}{{\tt arXiv:hep-th/0309177}}.

\bibitem{Henningson:1998cd}
M.~Henningson and K.~Sfetsos, ``{Spinors and the AdS/CFT correspondence},''
  \href{http://dx.doi.org/10.1016/S0370-2693(98)00559-0}{{\em Phys. Lett.} {\bf
  B431} (1998)  63--68},
\href{http://arxiv.org/abs/hep-th/9803251}{{\tt arXiv:hep-th/9803251}}.

\bibitem{Mueck:1998iz}
W.~Mueck and K.~S. Viswanathan, ``{Conformal field theory correlators from
  classical field theory on anti-de Sitter space. II: Vector and spinor
  fields},'' \href{http://dx.doi.org/10.1103/PhysRevD.58.106006}{{\em Phys.
  Rev.} {\bf D58} (1998)  106006},
\href{http://arxiv.org/abs/hep-th/9805145}{{\tt arXiv:hep-th/9805145}}.

\bibitem{Amsel:2008iz}
A.~J. Amsel and D.~Marolf, ``{Supersymmetric Multi-trace Boundary Conditions in
  AdS},'' \href{http://dx.doi.org/10.1088/0264-9381/26/2/025010}{{\em Class.
  Quant. Grav.} {\bf 26} (2009)  025010},
\href{http://arxiv.org/abs/0808.2184}{{\tt arXiv:0808.2184 [hep-th]}}.

\bibitem{Iqbal:2009fd}
N.~Iqbal and H.~Liu, ``{Real-time response in AdS/CFT with application to
  spinors},'' \href{http://dx.doi.org/10.1002/prop.200900057}{{\em Fortsch.
  Phys.} {\bf 57} (2009)  367--384},
\href{http://arxiv.org/abs/0903.2596}{{\tt arXiv:0903.2596 [hep-th]}}.

\bibitem{Beisert:2004ry}
  N.~Beisert,
  ``The dilatation operator of N = 4 super Yang-Mills theory and
  integrability,''
  {{\em Phys.\ Rept.}  {\bf 405} (2005) 1},
  \href{http://arxiv.org/abs/hep-th/0407277}{{\tt arXiv:hep-th/0407277}}.

\bibitem{Aharony:2003sx}
O.~Aharony, J.~Marsano, S.~Minwalla, K.~Papadodimas, and M.~Van~Raamsdonk,
  ``{The Hagedorn / deconfinement phase transition in weakly coupled large N
  gauge theories},'' {\em Adv. Theor. Math. Phys.} {\bf 8} (2004)  603--696,
\href{http://arxiv.org/abs/hep-th/0310285}{{\tt arXiv:hep-th/0310285}}.

\bibitem{Lucietti:2008cv}
J.~Lucietti and M.~Rangamani, ``{Asymptotic counting of BPS operators in
  superconformal field theories},''
  \href{http://dx.doi.org/10.1063/1.2970775}{{\em J. Math. Phys.} {\bf 49}
  (2008)  082301},
\href{http://arxiv.org/abs/0802.3015}{{\tt arXiv:0802.3015 [hep-th]}}.

\bibitem{Sundborg:1999ue}
B.~Sundborg, ``{The Hagedorn Transition, Deconfinement and N=4 SYM Theory},''
  \href{http://dx.doi.org/10.1016/S0550-3213(00)00044-4}{{\em Nucl. Phys.} {\bf
  B573} (2000)  349--363},
\href{http://arxiv.org/abs/hep-th/9908001}{{\tt arXiv:hep-th/9908001}}.

\bibitem{Polyakov:2001af}
A.~M. Polyakov, ``{Gauge fields and space-time},'' {\em Int. J. Mod. Phys.}
  {\bf A17S1} (2002)  119--136,
\href{http://arxiv.org/abs/hep-th/0110196}{{\tt arXiv:hep-th/0110196}}.

\bibitem{Maldacena:1997re}
  J.~M.~Maldacena,
  ``The large $N$ limit of superconformal field theories and supergravity,''
  {\em Adv.\ Theor.\ Math.\ Phys.\ } {\bf 2} (1998) 231
  [{\em Int.\ J.\ Theor.\ Phys.\ } {\bf 38} (1999) 1113],
\href{http://arxiv.org/abs/hep-th/9711200}{{\tt arXiv:hep-th/9711200}}.

\bibitem{Vafa:1994tf}
  C.~Vafa and E.~Witten,
  ``A Strong coupling test of S duality,''
  {\em Nucl.\ Phys.\  B} {\bf 431} (1994) 3,
  \href{http://arxiv.org/abs/hep-th/9408074}{{\tt arXiv:hep-th/9408074}}.

\bibitem{Polchinski:2000uf}
  J.~Polchinski and M.~J.~Strassler,
  ``The string dual of a confining four-dimensional gauge theory,''
  \href{http://arxiv.org/abs/hep-th/0003136}{{\tt arXiv:hep-th/0003136}}.

\bibitem{Callan:1997kz}
C.~G. Callan, Jr. and J.~M. Maldacena, ``Brane dynamics from the born-infeld
  action,'' {\em Nucl. Phys.} {\bf B513} (1998)  198--212,
\href{http://arxiv.org/abs/hep-th/9708147}{{\tt arXiv:hep-th/9708147}}.

\bibitem{Constable:1999ac}
N.~R. Constable, R.~C. Myers, and O.~Tafjord, ``{The noncommutative bion
  core},'' \href{http://dx.doi.org/10.1103/PhysRevD.61.106009}{{\em Phys. Rev.}
  {\bf D61} (2000)  106009},
\href{http://arxiv.org/abs/hep-th/9911136}{{\tt arXiv:hep-th/9911136}}.

\bibitem{DHoker:2007xy}
E.~D'Hoker, J.~Estes, and M.~Gutperle, ``{Exact half-BPS Type IIB interface
  solutions. I. Local solution and supersymmetric Janus},'' {\em JHEP} {\bf
  0706} (2007)  021, \href{http://arxiv.org/abs/arXiv:0705.0022}{{\tt
  arXiv:0705.0022 [hep-th]}}.

\bibitem{DHoker:2007xz}
E.~D'Hoker, J.~Estes, and M.~Gutperle, ``{Exact half-BPS Type IIB interface
  solutions. II. Flux solutions and multi-Janus},'' {\em JHEP} {\bf 0706}
  (2007)  022, \href{http://arxiv.org/abs/arXiv:0705.0024}{{\tt
  arXiv:0705.0024 [hep-th]}}.

\bibitem{Karch:2000gx}
A.~Karch and L.~Randall, ``{Open and closed string interpretation of SUSY CFT's
  on branes with boundaries},'' {\em JHEP} {\bf 06} (2001)  063,
\href{http://arxiv.org/abs/hep-th/0105132}{{\tt arXiv:hep-th/0105132}}.

\bibitem{Hawking:1982dh}
S.~W. Hawking and D.~N. Page, ``{Thermodynamics of Black Holes in anti-De
  Sitter Space},''
\href{http://dx.doi.org/10.1007/BF01208266}{{\em Commun. Math. Phys.} {\bf 87}
  (1983)  577}.

\bibitem{Witten:1998zw}
E.~Witten, ``{Anti-de Sitter space, thermal phase transition, and confinement
  in gauge theories},'' {\em Adv. Theor. Math. Phys.} {\bf 2} (1998)  505--532,
\href{http://arxiv.org/abs/hep-th/9803131}{{\tt arXiv:hep-th/9803131}}.

\bibitem{Berkooz:1998qp}
  M.~Berkooz and S.~J.~Rey,
  ``{Non-supersymmetric stable vacua of M-theory},''
  {\em JHEP} {\bf 9901} (1999) 014
  [{\em Phys.\ Lett.\  B} {\bf 449} (1999) 68],
\href{http://arxiv.org/abs/hep-th/9807200}{{\tt arXiv:hep-th/9807200}}.

\bibitem{Berkooz:2002ug}
M.~Berkooz, A.~Sever, and A.~Shomer, ``{'Double trace' deformations, boundary
  conditions and space-time singularities},'' {\em JHEP} {\bf 0205} (2002)
  034, \href{http://arxiv.org/abs/hep-th/0112264}{{\tt arXiv:hep-th/0112264
  [hep-th]}}.

\bibitem{Witten:2001ua}
E.~Witten, ``{Multitrace operators, boundary conditions, and AdS / CFT
  correspondence},'' \href{http://arxiv.org/abs/hep-th/0112258}{{\tt
  arXiv:hep-th/0112258 [hep-th]}}.

\bibitem{Gaiotto:2008sd}
D.~Gaiotto and E.~Witten, ``{Janus Configurations, Chern-Simons Couplings, And
  The theta-Angle in N=4 Super Yang-Mills Theory},''
  \href{http://dx.doi.org/10.1007/JHEP06(2010)097}{{\em JHEP} {\bf 1006} (2010)
   097}, \href{http://arxiv.org/abs/arXiv:0804.2907}{{\tt arXiv:0804.2907
  [hep-th]}}.

\bibitem{Gaiotto:2007qi}
D.~Gaiotto and X.~Yin, ``{Notes on superconformal Chern-Simons-Matter
  theories},'' \href{http://dx.doi.org/10.1088/1126-6708/2007/08/056}{{\em
  JHEP} {\bf 0708} (2007)  056},
  \href{http://arxiv.org/abs/arXiv:0704.3740}{{\tt arXiv:0704.3740
  [hep-th]}}.

\bibitem{Aharony:2008ug}
O.~Aharony, O.~Bergman, D.~L. Jafferis, and J.~Maldacena, ``{N=6 superconformal
  Chern-Simons-matter theories, M2-branes and their gravity duals},''
  {\em JHEP} {\bf 0810}, 091 (2008),
\href{http://arxiv.org/abs/0806.1218}{{\tt arXiv:0806.1218 [hep-th]}}.

\bibitem{Aharony:2008gk}
O.~Aharony, O.~Bergman, and D.~L. Jafferis, ``{Fractional M2-branes},''
{\em JHEP} {\bf 0811}, 043 (2008),
\href{http://arxiv.org/abs/0807.4924}{{\tt arXiv:0807.4924 [hep-th]}}.

\bibitem{Berman:2009kj}
D.~S. Berman and D.~C. Thompson, ``{Membranes with a boundary},''
  \href{http://dx.doi.org/10.1016/j.nuclphysb.2009.06.004}{{\em Nucl.Phys.}
  {\bf B820} (2009)  503--533},
  \href{http://arxiv.org/abs/arXiv:0904.0241}{{\tt arXiv:0904.0241
  [hep-th]}}.

\bibitem{Chu:2009ms}
C.-S. Chu and D.~J. Smith, ``{Multiple Self-Dual Strings on M5-Branes},''
  \href{http://dx.doi.org/10.1007/JHEP01(2010)001}{{\em JHEP} {\bf 1001} (2010)
   001}, \href{http://arxiv.org/abs/arXiv:0909.2333}{{\tt arXiv:0909.2333
  [hep-th]}}.

\bibitem{Berman:2009xd}
D.~S. Berman, M.~J. Perry, E.~Sezgin, and D.~C. Thompson, ``{Boundary
  Conditions for Interacting Membranes},''
  \href{http://dx.doi.org/10.1007/JHEP04(2010)025}{{\em JHEP} {\bf 1004} (2010)
   025}, \href{http://arxiv.org/abs/arXiv:0912.3504}{{\tt arXiv:0912.3504
  [hep-th]}}.
  
\bibitem{Chuetal}
C.-S. Chu and G.~S. Sehmbi, ``{Open M2-branes with flux and modified
Basu-Harcey equation},''
\href{http://arxiv.org/abs/arXiv:1011.5679}{{\tt arXiv:1011.5679
  [hep-th]}}.

\bibitem{abty}
O.~Aharony, M.~Berkooz, D.~Tong, and S.~Yankielowicz, work in progress.

\bibitem{Aharony:1999ti}
O.~Aharony, S.~S. Gubser, J.~M. Maldacena, H.~Ooguri, and Y.~Oz, ``{Large N
  field theories, string theory and gravity},''
  \href{http://dx.doi.org/10.1016/S0370-1573(99)00083-6}{{\em Phys. Rept.} {\bf
  323} (2000)  183--386},
\href{http://arxiv.org/abs/hep-th/9905111}{{\tt arXiv:hep-th/9905111}}.

\end{thebibliography}

\providecommand{\href}[2]{#2}\begingroup\raggedright\endgroup

\end{document}